\documentclass[12pt]{article}
\usepackage{amsmath,amssymb}
\usepackage[english]{babel}
\usepackage[cp866]{inputenc}
\usepackage{hyperref}

\oddsidemargin - 0.3 cm

\numberwithin{equation}{section}
\numberwithin{equation}{subsection}

\newcommand{\bq}{\begin{eqnarray}}
\newcommand{\eq}{\end{eqnarray}}
\newcommand{\bbq}{\begin{equation*}}
\newcommand{\eeq}{\end{equation*}}
\newcommand{\ra}{\rightarrow}
\newcommand{\ov}{\overline}

\newcommand{\la}{\Lambda_Q}

\newcommand{\lym}{\Lambda_{SYM}}

\newcommand{\mi}{m_{Q,i}}
\newcommand{\ml}{m_L}
\newcommand{\mh}{m_H}

\newcommand{\wt}{\widetilde}

\newcommand{\hh}{\mathbf{Q}^H_H}
\newcommand{\ohh}{\mathbf{{\ov Q}}^H_H}

\begin{document}

\begin{center} \vspace*{-7mm} {\bf \large  Mass spectra in ${\cal N}=1$ $\,SU(N_c)$ SQCD with $N_F=N_c$ \\ and problems with S-confinement} \end{center}

\begin{center}  \bf Victor L. Chernyak $^{\,a,\,b}$ \end{center}

\begin{center} \vspace*{-2mm} (e-mail: victorchernyak41@gmail.com) \end{center}

\begin{center} \vspace*{-2mm} $^a\,$ Novosibirsk State University, \,\, 630090 Novosibirsk, Pirogova str.2, Russia\end{center}

\begin{center} \vspace*{-2mm} $^b\,$ Budker Institute of Nuclear Physics SB RAS, \,\, 630090 Novosibirsk, Lavrentev ave.11, Russia\end{center}

\begin{center}  {\bf Abstract} \end{center}

The low energy mass spectra of the direct standard $SU(N_c),\,\,\,\,N_F=N_c$  $\,\,\,{\cal N}=1$ SQCD theory and its \,\, N. Seiberg's dual variant \cite{S1}, considered as {\it two independent theories}, are calculated in sections 2 and 3. It is shown that these two mass spectra are parametrically different,  both for equal or unequal small quark masses $0 < m_{Q,i}\ll\la$. Therefore, the proposal by N. Seiberg \cite{S1} of his $N_F=N_c$ dual theory of $N_F^2-1$ mesons $M^i_j$ and two baryons $B,\,{\ov B}$ as the low energy form of the direct theory is erroneous (and similarly for the case $N_F=N_c+1$, see Appendix {\bf B} in \cite{Session}).

In \cite{FS} the very special non-supersymmetric lattice $SU(N_c)$ QCD theory with $N_F=N_c$ scalar defective "quarks" in the unitary
gauge: \, $\Phi^i_{\beta}= \delta^i_{\beta}(|v|={\rm const} > 0),\,\,i,{\beta}=1,...,N_F=N_c$, was considered by E. Fradkin and S.H. Shenker. The conclusion of \cite{FS} was that the transition between the confinement at $0 < |v|\ll\Lambda_{QCD}$ (according to \cite{FS}) and higgs at $|v|\gg\Lambda_{QCD}$ regimes in this theory is the analytic crossover, not the non-analytic phase transition. And although the theory considered in \cite{FS} was very specific, the experience shows that up to now there is a widely spread opinion that this conclusion has general applicability.

This model used in \cite{FS} is criticized in section 5 of this paper as incompatible with and qualitatively different from the standard  not SUSY $SU(N_c)$ $\,\,N_F=N_c$ QCD theory with standard scalar quarks $\phi^i_{\beta}$ with all $2 N_c^2$ their physical real degrees of freedom. It is emphasized that this model \cite{FS} is really the Stueckelberg  $SU(N_c)$ YM-theory with no dynamical electric quarks and with
massive all $N_c^2-1$ electric gluons with fixed by hands nonzero masses. There is no genuine confinement in this theory, it stays permanently in the completely higgsed phase only. And this is a reason for a crossover in this theory. While in the theory with standard scalar quarks
there is the phase transition between the confinement and higgs phases.

In addition, in section 5, the arguments presented in \cite{IS} by K. Intriligator and N. Seiberg for the standard direct  $SU(N_c)$, $\,\,N_F=N_c\,\,\, {\cal N}=1$ SQCD in support of the crossover from \cite{FS} are criticized as erroneous.

Besides, in section 5 a qualitative mechanism of confinement in our ordinary (i.e. not SUSY) QCD is
proposed.

\vspace*{-6mm}

\tableofcontents
\numberwithin{equation}{section}

\section{Introduction}

\hspace*{4mm} N. Seiberg proposed in \cite{S1} the low energy form of the direct $SU(N_c)$\, ${\cal N}=1$ SUSY QCD theory (SQCD) with $N_F=N_c$ light quark flavors. While the high energy direct theory contained $2 N_F N_c$ colored complex quark fields $Q^i_{\beta},\,{\ov Q}^{\alpha}_i\,\,\, \beta=1...N_c,\,\, i=1...N_F=N_c$ and $N^2_c-1$ gluons, see \eqref{(2.2)}, the proposed in \cite{S1} low energy one at the scale  $\mu < \la$
($\la={\rm const}$ is the scale factor of the direct theory gauge coupling in the UV region at $\mi\ra 0$) contained instead only $N_F^2$ colorless complex mesons  $M^i_j=\sum_{\beta=1}^{N_c}({\ov Q}^{\beta}_j Q^i_{\beta}),\,\,i,j=1...N_F$ and two colorless complex baryons $B\sim Q^{N_c},\, {\ov B}\sim {\ov Q}^{\,N_c}$, with one constraint.

This regime with e.g. equal quark masses $m_{Q,i}=m_Q\ra 0$ and $\langle M^i_j\rangle=\delta^{i}_{j} \la^2,\,\, i,j=1...N_F$, see \eqref{(2.4)},  was called {\it "confinement with (spontaneous) chiral symmetry breaking"}, see e.g. section 4.2 in \cite{IS}. The implied physical interpretation was the following. At the scale $\mu \sim \la$ the gauge coupling is sufficiently large: $a(\mu=\la)=N_c g^2(\mu=\la)/8\pi^2={\cal O}(1)$. And  {\it all massless or light quarks and all gluons are confined, with the implied large string tension} $\sigma^{1/2}\sim\la$. They form then hadrons most of which have masses $\sim\la$ due to such string tension. And, at $m_Q\ll\la$, only special composites: $N_c^2-1$ independent mesons from $M^i_j=\sum_{\beta=1}^{N_c}
({\ov Q}^{\beta}_j Q^i_{\beta}),\,\,i,j=1...N_F=N_c$ and two baryons  $B\sim Q^{N_c}$ and ${\ov B}\sim {\ov Q}^{N_c}$  are light, with masses $\sim m_Q\ll\la$. For these reasons, only these enter the low energy Lagrangian of the direct theory at the scale $\mu < \la$. As a test, it was checked that the 't Hooft triangles are matched, see \cite{S1}. This Seiberg's proposal is usually called "S-confinement"\, in the literature.\\

The above line of reasonings was criticized in \cite{ch1}, see section 7 therein.
\footnote{\,
It was pointed out in \cite{ch1} that {\it the confinement originates only from} SYM. There is no confinement in Yukawa-type theories without gauge interactions. But the lower energy SYM theory contains only one universal dimensional parameter $\lym=\Bigl(\la^{3N_c-N_F}\Pi_{i=1}^{N_F}\mi\Bigr )^{1/3 N_c}\ll\la$ at $\mi\ll\la$,\, see \eqref{(2.3)}. So that, {\it its string tension at $\mi\ll\la$ can not be as large as $\sigma^{1/2}\sim\la$, but is much smaller:\, $\sigma^{1/2}\sim\lym\ll\la$. And $\lym\ra 0$ if even one $m_{Q,i}\ra 0$, in which limit there is no confinement at all}. \label{(f1)}
}
And this argument from footnote \ref{(f1)} is applicable  to all values $N_F > 0$ at all $\mi\ll\la$. Therefore, {\it the interpretation e.g. of the dual theory \eqref{(3.1.3)},\eqref{(3.1.4)} with $\mi=m_Q$ as the low energy form of the direct theory \eqref{(2.2)} at $m_Q\ll\la$ due to strong confinement with $\sigma^{1/2}\sim\la$ is not correct. It should be understood as independent theory. And mass spectra of these two theories have to be calculated and compared to see whether they are equivalent or not}.

And similarly for the case $N_F=N_c+1$. The theory with $N_F^2$ colorless mesons  $M^i_j=\sum_{\beta=1}^{N_c}({\ov Q}^{\beta}_j Q^i_{\beta})$ and $2 N_F$ colorless baryons $B^i\sim Q^{N_c+1},\,\, {\ov B}\sim {\ov Q}^{N_c+1}$ was proposed by N. Seiberg as the low energy form of this direct theory and similarly interpreted in \cite{S1},\cite{IS} as the regime with "confinement without (spontaneous) chiral symmetry breaking"\, at e.g. $\mi=m_Q\ra 0$.\\

In \cite{S2} N.Seiberg introduced the dual "magnetic"\, $SU({\ov N}_c=N_F-N_c)$ gauge theory with e.g. $N_c+2\leq N_F < 3N_c$ dual quarks $q^a_i,\,\,
{\ov q}^i_a,\,\,i=1...N_F, \,\,a=1...{\ov N}_c$. And proposed the following. -  a) {\it "The quarks and gluons of one theory can be interpreted as solitons (non-Abelian magnetic monopoles) of the elementary fields of the other theory}\, \cite{S2}"\,.\, b) These two theories, although clearly different at scales $\mu > \la$, {\it become equivalent} at $\mu < \la$. I.e., {\it the dual theory is the equivalent alternative description of the direct one} at $\mu < \la$. {\it "When one of the theories is Higgsed by an expectation value of a squark, the other theory is confined. Massless glueballs, baryons and Abelian magnetic monopoles in the confining description are the weakly coupled elementary quarks (i.e. solitons of the confined quarks) in the dual Higgs description}\, \cite{S2}"\,.

The Seiberg duality \cite{S1},\cite{S2} passed a number non-trivial checks, mainly in the effectively massless regime: the matching of 't Hooft triangles and behavior in conformal regimes, and correspondences under appropriate mass deformations. But up to now no direct proof has been given that the direct and dual theories are really equivalent. The reason is that such a proof needs real understanding of and the control over the dynamics of both theories in the strong coupling regimes. E.g., at $N_c+2 \leq N_F < 3N_c$, no way was shown up to now to obtain solitons with quantum numbers of dual quarks in the direct theory (and vice versa). Therefore, the Seiberg proposal of equivalence of direct and dual theories at $\mu < \la$ remains a hypothesis up to now, see e.g. Introduction in arXiv:0811.4283 [hep-th] (\cite{ch3}) and references \cite{AR,G,A,GK,V}.\\

For the reasons described above, our approach is to consider the direct and Seiberg's dual theories {\it as independent theories,  to calculate their mass spectra and to compare}. This is then the direct {\it additional check} of possible equivalence of these two theories at $\mu < \la$.

For this purpose, the dynamical scenario was introduced in \cite{ch3} which allows to calculate the mass spectra in ${\cal N}=1$ SQCD-type theories even in strong coupling regimes. The only dynamical assumption of this scenario  is that quarks (with $\mi\neq 0$) in such theories can be in two different {\it standard} phases only\,: this is either the HQ (heavy quark) phase where they are not higgsed but confined at $\lym\neq 0$, or the higgs phase where they form nonzero coherent condensate in the vacuum breaking the color symmetry, at the appropriate values of the Lagrangian parameters. The word "standard" implies also that, in addition to the ordinary mass spectrum described e.g. in \cite{ch3},\cite{ch5} and  below in this paper, in such ${\cal N}=1$ theories without elementary colored adjoint scalars (unlike the very special ${\cal N}=2$ SQCD with its additional elementary colored scalar fields $X^{adj}$ and enhanced supersymmetry) no {\it additional} parametrically lighter solitons (e.g. magnetic monopoles or dyons) are formed at those scales where the {effectively} massless regime is broken explicitly by nonzero masses  $\mi\neq 0$ of not higgsed quarks in the HQ-phase. (It is worth noting that the appearance of {\it additional} light solitons will influence the  't Hooft triangles of this  ${\cal N}=1$ theory). Let us emphasize also that this dynamical scenario satisfies all those tests which were used as checks of the Seiberg hypothesis about the equivalence of the direct and dual theories at scales $\mu < \la$. This shows, in particular, that all these tests, although necessary, may well be insufficient.

Within the framework of this dynamical scenario, in was shown e.g. in \cite{ch3},\cite{ch5} and in Appendix {\bf B} of \cite{Session} that mass spectra of the direct theories and their Seiberg's dual variants are parametrically different in many cases under control. Therefore, {\it the Seiberg dual $SU({\ov N}_c=N_F-N_c)$ theory is not the genuine or equivalent low energy form of the direct theory at $\mu < \la$. It is the independent theory}. And, in each case, the mass spectra of the direct and dual theories should be calculated and compared to see whether they are really equivalent or not. This is a purpose of this article for the case $N_F=N_c$.

The organization of this paper is seen from the contents.

\section{Direct theory with $N_F=N_c$}.

Let  us start with the direct theory at $\mu=\la$ with $N_c$ colors and  $N_F=N_c+1$ quark flavors with e.g. {\it the real positive masses}:
$0 < \mi(\mu=\la)=m_{Q,i}\ll\la,\,\,i=1...N_c,\,\, m_{Q,N_c+1}(\mu=\la)=\la$. The Lagrangian at $\mu=\la$ looks as
\footnote{\,
The gluon exponents are implied in Kahler terms. \label{(f2)}
}
\bq
K_{N_F=N_c+1}={\rm Tr}_{N_F=N_c+1}\Bigl (Q^\dagger Q+Q\ra {\ov Q} \Bigr ),\quad {\cal W}_{N_F=N_c+1}=-\frac{2\pi}{\alpha(\mu=\la)} S+\sum_{i=1}^
{N_F=N_c}\mi M^i_i +\la M^{N_c+1}_{N_c+1}\,,\,\, \label{(2.1)}
\eq
\bbq
M^i_j=\sum_{\beta=1}^{N_c}({\ov Q}^{\beta}_j Q^i_{\beta})(\mu=\la)\,, \quad  i,j=1...N_F=N_c\,, \quad M^{N_c+1}_{N_c+1}=\sum_{\beta=1}^{N_c}({\ov Q}^{\beta}_{i=N_c+1} Q^{i=N_c+1}_{\beta})(\mu=\la)\,,
\eeq
where $S=\sum_{A,\gamma} W^{A,\,\gamma} W^{A}_{\gamma}/32\pi^2$,\,\, $W^A_{\gamma}$ is the gauge field strength, $A=1...N_c^2-1,\, \gamma=1,2$,\, $a(\mu)=N_c g^2(\mu)/8\pi^2=N_c\alpha(\mu)/2\pi$ is the gauge coupling with its scale factor $\la={\rm const}$.

Integrating out the last heaviest quark, we obtain at $\mu=\la$ the Lagrangian of the direct theory with $N_F=N_c$ flavors of light quarks (and with the same scale factor $\la$ of the gauge coupling)
\bq
K_{N_F=N_c}={\rm Tr}_{N_F=N_c}\Bigl (Q^\dagger Q+Q\ra {\ov Q} \Bigr ),\quad {\cal W}_{N_F=N_c}=-\frac{2\pi}{\alpha(\mu=\la)} S+\sum_{i=1}^{N_F=N_c}\mi M^i_i\,.\label{(2.2)}
\eq

To find the mean vacuum value of the {\it colorless} gluino condensate $\langle S\rangle_{N_c}$ in this theory, we use that it depends analytically on the numerical values $\mi$ of quark masses. For this reason, we can start now with all very heavy quarks with $\mi\sim m_{Q,j}\gg \la$ in the HQ (heavy quark) phase, i.e. not higgsed but confined,  and integrate them out. There remains the pure $SU(N_c)$ SYM lower energy theory with the scale factor of its gauge coupling $\lym$ and the gluino mean vacuum value, see \cite{VY} and section 2 in \cite{ch1},
\bq
\lym^3=\langle S\rangle_{N_c}=\frac{\sum_{\gamma=1}^{2}\sum_{A=1}^{N_c^2-1}\langle \lambda^{A,\gamma}\lambda^A_{\gamma}\rangle}{32\pi^2}=
\Bigl (\la^{3N_c-N_F=2N_c}\Pi_{i=1}^{N_F=N_c}\mi\Bigr )^{1/N_c}. \label{(2.3)}
\eq
And now the values $\mi$ in \eqref{(2.3)} can be continued analytically to desired values, e.g. $0 < \mi\ll\la$.

From the Konishi anomalies \cite{Konishi}, see \eqref{(2.2)},
\bq
\langle M^i_j\rangle=\sum_{\beta=1}^{N_c}\langle {\ov Q}^{\beta}_j Q^i_{\beta}\rangle=\delta^i_j\frac{\langle S\rangle_{N_c}}{\mi}=\delta^i_j \la^2\frac{\det^{1/N_c} \mi\equiv \Bigl (\Pi_{i=1}^{N_F=N_c}\mi\Bigr )^{1/N_c}}{\mi}\,.\label{(2.4)}
\eq

Besides, from \eqref{(2.4)}
\bq
\langle \det M^i_j\rangle=\Pi_{i=1}^{N_c}\langle M^i_i\rangle=\frac{\Bigl (\langle S\rangle_{N_c}\Bigr )^{N_c}}{\det m_{Q,i}} = \la^{2N_c}\,, \label{(2.5)}
\eq
{\it for all  $\mi \neq 0$, and even in the chiral limit $\mi\,\ra\,0$}. In this limit, the theory \eqref{(2.2)} approaches to the moduli space. \\

Let us consider now the colorless baryon operator
\bq
B=\frac{1}{N_c !\,\la^{N_c-1}}\det Q^i_\beta=\frac{1}{N_c !\,\la^{N_c-1}}\sum_{i=1}^{N_F=N_c}\sum_{\beta=1}^{N_c}\epsilon_{i_1...i_{N_c}}
\epsilon^{\beta_1...\beta_{N_c}} Q^{i_1}_{\beta_1}...Q^{i_{N_c}}_{\beta_{N_c}}, \label{(2.6)}
\eq
and similarly ${\ov B}$ with $Q\ra {\ov Q}$.  Let us start first with heavy equal mass quarks with $\mi=m_Q\gg\la$. If all such quarks were higgsed
then, by definition, $\langle B\rangle=\langle {\hat Q}^1_1\rangle...\langle{\hat Q}^{N_c}_{N_c}\rangle\neq 0$. Here $\langle{\hat Q}^i_{\beta}
\rangle=\delta^i_{\beta}\rho_{\rm higgs}$, where $\rho_{\rm higgs}\neq 0$ is the gauge invariant order parameter introduced in section 2 of \cite{ch21}. But, clearly, all such heavy quarks are in the HQ (heavy quark) phase, i.e. {\it not higgsed but confined} by $SU(N_c)$ SYM\,:\, $\langle {\hat{\ov Q}}_i^{\beta}\rangle=\delta_i^{\beta}\rho_{HQ}=0,\,\, \langle {\hat Q}^i_{\beta}\rangle=\delta^i_{\beta}\rho_{HQ}=0,\,\,\,  i,\beta=1...N_c$, see section 4.1 in \cite{ch21}. Therefore, $\langle B\rangle=\langle{\ov B}\rangle=0$ for such equal mass heavy quarks. While $\langle M^1_1\rangle=\langle M^i_i\rangle=\la^2\neq \langle {\hat Q}^1_1\rangle
\langle{Q}^{1}_{1}\rangle=0$ for such heavy equal mass quarks, see \eqref{(2.4)}. The reason is that $\langle M^i_i\rangle$ is non-factorizable in this case and is nonzero due to the Konishi anomaly.
\footnote{\,
There is a qualitative difference between $\langle B\rangle=\langle{\ov B}\rangle$ and $\langle M^i_i\rangle$. This last is non-factorizable in the HQ-phase with $\rho_{HQ}=0$, but is nonzero even for heavy non-higgsed quarks in this HQ-phase due to the one quark loop Konishi anomaly, \eqref{(2.4)}, see section 4.1 in \cite{ch21}. But there is no analog of the Konishi anomaly for  $\langle B\rangle=\langle{\ov B}\rangle$.  \label{(f3)}
}

But, as mean vacuum values of lowest components of any {\it colorless} chiral superfields, e.g. colorless $\langle{\rm Tr}\, \lambda\lambda\rangle$ or $\langle M^i_i\rangle$, the mean vacuum values $\langle B\rangle=\langle{\ov B}\rangle$ of these {\it colorless} chiral superfields also {\it depend analytically on values of the chiral superpotential parameters $\mi$}. Therefore, for these equal mass quarks, the values of $\langle B\rangle =\langle{\ov B}\rangle$ can be continued analytically from the regions of heavy quark mass values where definitely $\langle B\rangle=\langle {\ov B}\rangle=0$  to {\it small nonzero mass values}, and both $\langle B\rangle$ and $\langle{\ov B}\rangle$ will remain equal zero.

In particular, it follows from the above that {\it even all light equal mass quarks are in the HQ-phase and not higgsed}
\bq
\langle {\hat Q}^i_{\beta}\rangle=\langle{\hat {\ov Q}}^{\beta}_i\rangle=0,\,\,\, i,\beta=1... N_F=N_c,\,\,\, {\rm for}\,\,\, 0 < \mi=m_Q\ll\la,\,\, {\rm and\,\,\,even\,\,\,in\,\,\,the\,\,\,limit}\,\, m_Q\ra 0\,.  \label{(2.7)}
\eq

As it is seen from \eqref{(2.4)},\eqref{(2.5)},\eqref{(2.2.1)}, it is impossible to have all $\langle M^i_i\rangle > \la^2$ for all $N_F=N_c$ unequal mass quarks. Heavier H-quarks will have $\langle M^H_H\rangle < \la^2$. And because these quarks were not higgsed even for equal mass quarks with  $\langle M^H_H\rangle = \la^2$ \eqref{(2.4)},\eqref{(2.5)},\eqref{(2.7)}, they definitely will remain non-higgsed at $\langle M^H_H\rangle < \la^2$, see \eqref{(2.2.1)}. But when even one H-quark is not higgsed,  then $\langle B\rangle=\Pi_{i=1}^{N_F=N_c}\langle Q^i_{i}\rangle=0$ also for unequal mass quarks with $m_{Q,i} > 0$, and  even in the chiral limit $m_{L,H}\ra 0$ with fixed ratio $0 < r=m_L/m_H\ll 1$, see section 2.2, footnote \ref{(f3)} and \eqref{(4.1)}. And the same for $\langle{\ov B}\rangle=\langle B\rangle=0$.

\numberwithin{equation}{subsection}
\subsection{Light equal mass  quarks}

Let us consider first just this case with all light equal mass quarks, $0 < \mi=m_Q\ll\la\,,\,\,\,i=1...N_F=N_c$. As shown above, see \eqref{(2.7)},
{\it they all are in the HQ (heavy quark) phase, i.e. not higgsed but confined} by $SU(N_c)$ SYM. Lowering the scale $\mu$ from $\mu=\la$ down to $\mu=m^{\rm pole}_{Q}\ll\la$,
the Lagrangian at $\mu=m^{\rm pole}_{Q}$ has the form
\bq
K_{N_F=N_c}=z_Q(\la,\mu=m^{\rm pole}_{Q}){\rm Tr}_{N_F=N_c}\Bigl (Q^\dagger Q+Q\ra {\ov Q} \Bigr )\,, \label{(2.1.1)}
\eq
\bbq
{\cal W}_{N_F=N_c}=-\frac{2\pi}{\alpha(\mu=m^{\rm pole}_{Q})} S+ m_Q \sum_{i=1}^{N_c} M^i_i\,,
\eeq
\bbq
z_Q(\la,\mu=m^{\rm pole}_{Q})\sim\Bigl (\frac{\mu=m^{\rm pole}_{Q}}{\la}\Bigr )^{\gamma^{\rm str}_Q}\ll 1,\quad m^{\rm pole}_{Q}=\frac{m_Q}{z_Q(\la,\mu=m^{\rm pole}_{Q})}\sim \la \Bigl (\frac{m_{Q}}{\la} \Bigr )^{0\, <\, \frac{1}{1+\gamma^{\rm str}_Q}\, < \,\frac{1}{3}}\ll\la\,,
\eeq
where $\gamma^{\rm str}_Q$ is the quark anomalous dimension, see \eqref{(2.1.2)}.

As explained in section 7 of \cite{ch1} or in section 3 of \cite{ch21}, in the considered UV-free ${\cal N}=1$ SQCD with $N_F=N_c$, if quarks are not higgsed at scales $\mu > \la$, see \eqref{(2.7)}, then this theory enters smoothly at $\mu < \la$ into the perturbative strong coupling regime with the gauge coupling $a(\mu\ll\la)=N_c g^2(\mu\ll\la)/8\pi^2\gg 1$ and with effectively massless all quarks and gluons at $m^{\rm pole}_{Q} < \mu < \la$.  The perturbatively exact  NSVZ $\beta$ function \cite{NSVZ} looks then as
\bq
\frac{d a(\mu)}{d\ln \mu} = \beta(a)=-\, \frac{a^2}{1-a}\, \frac{(3N_c-N_F)-N_F\gamma^{\rm str}_Q(a)} {N_c}\,\,\, \ra \,\,\, -\, \nu\, a\,<\, 0,
\label{(2.1.2)}
\eq
\bbq
\nu=\Bigl [\frac{N_F}{N_c}(1+\gamma^{\rm str}_Q) - 3\Bigr ]=\Bigl [\gamma^{\rm str}_Q-2\Bigr ]={\rm const} > 0\,, \quad
a(\mu\ll\la)\sim\Bigl (\frac{\la}{\mu} \Bigr )^{\nu\, >\, 0}\gg 1\,.
\eeq
In section 7 of \cite{ch1} the values $\gamma^{\rm str}_Q=(2N_c-N_F)/(N_F-N_c) > 1,\,\,\nu=(3N_c-2N_F)/(N_F-N_c) > 0$ at $\mu\ll\la$ and $N_c < N_F < 3N_c/2$ have been found from matching of definite two point correlators in the direct  $SU(N_c)$ theory and in Seiberg's dual \cite{S2}. But this is not applicable in our case here with $N_F = N_c$.  So that, unfortunately, we can't find the concrete value $\gamma^{\rm str}_Q$. But, as will be shown below, for our purposes it will be sufficient to have the only condition $\nu > 0$ in \eqref{(2.1.2)}.

At $\mu= m^{\rm pole}_Q$ all quarks decouple as heavy and there remains at lower energies $SU(N_c)$ SYM in the perturbative strong coupling branch. From the
NSVZ $\beta$-function \cite{NSVZ}
\bbq
\frac{d a^{(str,\, pert)}_{SYM}(\mu\gg\lym)}{d\ln\mu}= -\,\, \frac{3\,\Bigl (a^{(str,\, pert)}_{SYM}(\mu\gg\lym)\Bigr ) ^2}{1-a^{(str,\, pert)}_{SYM}(\mu\gg\lym)}\ra 3\, a^{(str,\, pert)}_{SYM}(\mu)\,,
\eeq
\bq
a^{(str,\, pert)}_{SYM}(\mu\gg\lym)\sim\Bigl (\frac{\mu}{\lym}\Bigr )^3 \gg 1\,, \quad a^{(str,\, pert)}_{SYM}(\mu\sim\lym)={\cal O}(1)\,. \label{(2.1.3)}
\eq
The scale factor $\lym$ of the gauge coupling is determined from matching, see \eqref{(2.1.1)}-\eqref{(2.1.3)}
\bq
a_{+}(\mu=m^{\rm pole}_Q)=\Bigl (\frac{\la}{m^{\rm pole}_Q} \Bigr )^{\nu}=a^{(str,\, pert)}_{SYM}(\mu=m^{\rm pole}_Q)=\Bigl (\frac{m^{\rm pole}_Q)}{\lym} \Bigr )^3\, \ra  \label{(2.1.4)}
\eq
\bbq
\ra \,\,\lym^3=\Bigl (\la^{2N_c} m^{N_c}_Q \Bigr )^{1/N_c}=m_Q\la^2,\,\,\,
\eeq
as it should be, see \eqref{(2.3)}. Besides,  as a check of self-consistency, see \eqref{(2.1.1)},\eqref{(2.1.2)},\eqref{(2.1.4)}
\bq
\Bigl (\frac{\lym}{m^{\rm pole}_Q}\Bigr )^3\sim\Bigl (\frac{m_Q}{\la} \Bigr )^{\omega\, >\, 0}\ll 1\,,\quad \omega=
\frac{\nu > 0}{(1+\gamma^{\rm str}_Q)} > 0 \,,  \label{(2.1.5)}
\eq
as it should be.  At $\mu < \lym$ the perturbative RG evolution stops due to nonperturbative effects $\sim\lym$ in the pure ${\cal N}=1\,\,\,SU(N_c)$ SYM.\\

After integrating {\it inclusively} all quarks as heavy at $\mu < m^{\rm pole}_{Q}$ and then all gluons at $\mu < \lym$ via the VY procedure \cite{VY}, the low energy superpotential at $\mu < \lym$
\bq
{\cal W}=N_c\Bigl (\la^{2N_c} m^{N_c}_Q \Bigr )^{1/N_c}=N_c m_Q\la^2=N_c \lym^3 \label{(2.1.6)}
\eq
looks as it should be.

Therefore, the mass spectrum of the direct theory for this case looks as follows. a)\,All quarks are in the HQ (heavy quark) phase. i.e. {\it not higgsed but confined}, and decouple as heavy at $\mu < m^{\rm pole}_Q$, see \eqref{(2.1.1)}.\\
b)\, There is a number of quarkonia with the typical mass scale ${\cal O}(m^{\rm pole}_Q)\gg \lym$. {\it This mass scale is checked by eqs.\eqref{(2.1.4)},\eqref{(2.1.5)}. Integrating inclusively all these quarkonia  (i.e. equivalently, integrating inclusively all quarks as heavy at $\mu= m^{\rm pole}_Q$), we obtain the well known beforehand right value of $\lym$}, see \eqref{(2.3)}.\\
c)\, there is a number of $SU(N_c)$ gluonia with the typical mass scale ${\cal O}(\lym)$ (here and below: at least at not too large $N_c$).

In the chiral limit $m_Q\ra 0$, the masses of {\it all} quarks, gluons and hadrons  {\it coalesce to zero}. Because the tension of the $SU(N_c)$ confining string is $\sigma^{1/2}\sim\lym\ra 0$, {\it there is no confinement} in this chiral limit, see Introduction and footnotes \ref{(f1)},\,\ref{(f4)}.

For more details see section 4.

\numberwithin{equation}{subsection}
\subsection{Light unequal mass  quarks}

Consider now unequal quark masses \,: $\mi=\ml,\,\,i=1,...,N_L,\,\,\,m_{Q,i}=\mh,\,\, i=N_L+1,...,N_F ,\,\,\,N_L+N_H=N_F=N_c$, with $0 < m_L\ll m_H\ll\la$. In this case, see \eqref{(2.3)}-\eqref{(2.5)},
\bq
\langle M_L\rangle=\langle M^L_L\rangle=\frac{\langle S\rangle_{N_c}}{\ml}=\la^2\Bigl ( \frac{1}{r}\Bigr )^{\frac{N_H}{N_c}},\,\,
\langle M_H\rangle=\langle M^H_H\rangle=\frac{\langle S\rangle_{N_c}}{\mh}=\la^2\,r^{\frac{N_L}{N_c}},\,\, \langle M_L\rangle\gg\la^2\,,\,\, \langle M_H\rangle\ll\la^2.\,\,\,\,\,\,\, \label{(2.2.1)}
\eq
\bbq
\langle M_L\rangle^{N_L}\langle M_H\rangle^{N_H}=\la^{2 N_c},\,\,  0 < r=\ml/\mh\ll 1,\,\, \langle B\rangle= \langle{\ov B}\rangle=0\,.
\eeq
\bbq
(M_{L,H})^i_j=(M_{L,H})^i_j(\rm adj)+ \delta^i_j\, M_{L,H}({\rm singl}),\,\, M_{L,H}({\rm singl})=\frac{1}{N_{N_L,N_H}}\sum_{i=1}^{N_L,N_H} M^i_i\,\,,
\eeq
\bbq
\langle (M_{L,H})^i_j(adj)\rangle=0, \,\, \langle M_{L,H}({\rm singl})\rangle=\langle M_{L,H}\rangle\,.
\eeq

In this case, at not too large $N_c$, see \cite{ch21} and footnote \ref{(f5)}, \, $Q^L,\,{\ov Q}_L$ quarks are higgsed at the large scale $\mu=\mu_{\rm gl,L}\gg\la$ in the weak coupling region. There are $N_L(2N_c-N_L)$ heavy gluons (and their superpartners) with masses (ignoring logarithmic renormalization factors),
\bq
\frac{\mu^2_{\rm gl,\, L}}{\la^2}\sim \frac{1}{N_c}\frac{\rho^2_{\rm higgs}}{\la^2}\sim \frac{1}{N_c}\frac{\langle M_L({\rm singl})\rangle}{\la^2}\sim
\frac{1}{N_c}\frac{\langle S\rangle_{N_c}}{m_L\la^2}\sim\frac{1}{N_c}\Bigl ( \frac{1}{r}\Bigr )^{\frac{N_H}{N_c}}\gg 1\,,\label{(2.2.2)}
\eq
\bbq
\langle {\hat Q}^i_{\beta}(x)\rangle =\langle {\hat Q}^i_{\beta}(0)\rangle=\delta^i_{\beta}\,\rho_{\rm higgs}\,, \quad \frac{\rho^2_{\rm higgs}=
\langle M_L({\rm singl})\rangle}{\la^2}=\Bigl ( \frac{1}{r}\Bigr )^{\frac{N_H}{N_c}} \gg 1\,,\quad i=1...N_L\,,\,\,\,\beta=1...N_c\,,
\eeq
where ${\hat Q}^i_{\beta}(x)$ is the {\it colored} gauge invariant quark field and $\rho_{\rm higgs}$ is the gauge invariant order parameter, see section 2 in \cite{ch21}.

In the chiral limit $m_{L,H}\ra 0$ with fixed ratio $0 < r=m_L/m_H \ll 1$, the Lagrangian \eqref{(2.2)} is invariant under global $SU(N_{F})_{left}\times SU(N_{F})_{right}$ flavor transformations under which quarks transform in the fundamental and anti-fundamental representations:\, $Q^{left}_{\beta},\,
{\ov Q}^{\beta}_{right}$. The generators of global V- or A- flavor transformations are, respectively, $(right+left)$ and $(right-left)$. The upper and lower flavor indices denote fundamental and anti-fundamental representations. Besides, in what follows, e.g. $Q^{L,H}_{\beta}$ and ${\ov Q}^{\beta}_{L,H}$ will denote left $Q$ quarks with flavor indices $L=(i=1,...,N_L)$ or $H=(i=N_L+1,...,N_F)$, and similarly for right $\ov Q$ quarks. And similarly for the color indices:\, $L=(\beta=1,...,N_L)$ or $H=(\beta=N_L+1,...,N_c)$, e.g. $Q^L_L$.

At $1\leq N_L\leq N_c-2$ and small nonzero $m_{L,H}$ there remains at lower energies $\mu < \mu_{\rm gl.L}$\,:\, $SU(N^\prime_c=N_c-N_L=N_H\geq 2)$ gauge group with the scale factor $\Lambda_H$ of its gauge coupling, $N^2_L$ colorless genuine complex pseudo-Goldstone pions $\pi^L_L(adj)$ (sitting on $\Lambda_H$) remained from higgsed $Q^L_L,\, {\ov Q}^L_L$ quarks, $2N_L N_H$ equal mass hybrid genuine complex pseudo-Goldstone pions $\pi^H_L,\,\pi^L_H$ (in essence, these are non-active $Q^H_L,\, {\ov Q}^L_H$ quarks with higgsed L-colors), and $2 N^2_H$  of still active non-higgsed but confined $\hh,\,\ohh$ quarks with $N_H$ flavors and $N_H$ non-higgsed colors.

The fields of hybrid pions canonically normalized at $\mu=\la$ are defined as
\bq
\pi^H_L=\frac{1}{\langle M_L\rangle^{1/2}}\sum_{\beta=1}^{N_L}\langle{\ov Q}_L^{\beta}\rangle Q^H_{\beta},\quad \pi^L_H=\frac{1}{\langle M_L\rangle^{1/2}}
\sum_{\beta=1}^{N_L}{\ov Q}^{\beta}_H\langle  Q^L_{\beta}\rangle,\quad H=(N_L+1,...,N_F),\quad \beta=(1,...,N_L).\,\,\,\, \label{(2.2.3)}
\eq

After all heavy particles with masses $\sim \mu_{\rm gl,L}$ decoupled at $\mu < \mu_{\rm gl,L}$, the scale factor $\Lambda_H$ of the remained $SU(N_H)$ gauge
theory is
\bq
\langle\Lambda_H^{3 N^\prime_c-N^\prime_F=2 N_H}\rangle=\frac{\la^{2 N_c}}{\langle M_L\rangle^{N_L}}\,,\quad \langle\Lambda_H^2\rangle=
\la^2 r^{N_L/N_c}=\langle M_H\rangle \ll \la^2\,,\quad \Lambda^{2N_H}_H=\frac{\la^{2 N_c}}{\det{M^L_L}}\,.  \label{(2.2.4)}
\eq

The heavy hybrid gluons $(A_{\mu})^{L}_{H},\,\,(A_{\mu})_{L}^{H}$ with masses $\mu_{\rm gl,L}\gg\la$ and lighter active H-quarks $\hh,\,\ohh$ with $N_H\geq 2$ non-higgsed colors are {\it weakly confined} by $SU(N_H)$ SYM (the string tension is $\sigma^{1/2}\sim\langle\lym\rangle$, much smaller than their perturbative pole masses). These hybrid gluons decouple as heavy at $\mu < \mu_{\rm gl,L}$. Quarks $\hh,\,\ohh$ are in the  HQ (heavy quark) phase, i.e. not higgsed but confined and decouple at $\mu < m^{\rm pole}_H\ll\la\ll \mu_{\rm gl,L}$.

The Lagrangian of lighter particles  looks at $\mu= \mu_{\rm gl,L}$ as
\bq
K=2 z_Q(\la,\mu_{\rm gl,L}) {\rm Tr}\sqrt{(M^L_L)^\dagger M^L_L}+z_Q(\la,\mu_{\rm gl,L}){\rm Tr}\Bigl ( (\pi^H_L)^\dagger \pi^H_L+(\pi_H^L)^\dagger \pi_H^L\Bigr ) +  \label{(2.2.5)}
\eq
\bbq
 + z_Q(\la,\mu_{\rm gl,L}){\rm Tr}\Bigl ( (\hh)^\dagger \hh + ({\ohh})^\dagger \ohh \Bigr )\,,
\eeq
\bbq
{\cal W}={\cal W}_{SU(N_H)}^{SYM} +m_L{\rm Tr} M_L + m_H {\rm Tr}\Bigl ( \pi^H_L \pi^L_H + \ohh \hh \Bigr)\,,
\eeq
where $z_Q(\la,\mu_{\rm gl,L})\sim \log^{1/2} (\mu_{\rm gl,L}/\la)\gg 1$ is the logarithmic renormalization factor of quarks.

{\bf A)}\, If $\la\gg m_H \gg \langle\Lambda_H\rangle=\langle M_H\rangle^{1/2}$, see \eqref{(2.2.4)}\,: active HH-quarks $\hh,\, \ohh$ are in the HQ (heavy quark) phase and decouple as heavy at $\mu < m^{\rm pole}_H=m_H/z_Q(\la, m^{\rm pole}_H)\gg \langle\Lambda_H\rangle$ {\it in the weak coupling regime}, where
$z_Q(\la, m^{\rm pole}_H)\ll 1$ is the logarithmic renormalization factor of active HH-quarks.

There remain at lower energies: $SU(N_H)$ SYM theory, $N_L^2$ light genuine colorless complex $\pi^L_L$ pions \eqref{(2.2.8)} and light genuine hybrid complex pions $\pi^H_L, \pi^L_H$ \eqref{(2.2.3)}.

As a check, the  gluon mass due to potentially possible higgsing of active HH-quarks looks as ( with logarithmic accuracy), see \eqref{(2.2.4)},
\bq
m^{\rm pole}_H=\frac{m_H}{z_Q(\la, m^{\rm pole}_H)}\sim m_H\gg \langle\Lambda_H\rangle\gg\langle\lym\rangle,\,\, z_Q(\la, m^{\rm pole}_H)\sim \log^{-1/2} (\la/m^{\rm pole}_H)\ll 1\,, \label{(2.2.6)}
\eq
\bbq
\frac{\mu^2_{\rm gl,\,H}}{m^{2,\,\rm pole}_H}\sim\frac{\langle M_H\rangle=\langle\Lambda_H^2\rangle}{m^{2,\,\rm pole}_H\sim m^2_H} \ll 1\,,
\eeq
where $z_Q(\la, m^{\rm pole}_H)\ll 1$ is the quark perturbative logarithmic renormalization factor.

This shows that active HH-quarks $\hh,\, \ohh$ are in the HQ (heavy quark) phase, i.e. not higgsed but confined by $SU(N_H)$ SYM.\\

After integrating inclusively decoupled at $\mu < m^{\rm pole}_H$  active HH-quarks, the scale factor ${\hat\Lambda}_{N_H}$ of the gauge coupling of remained $SU(N_H)$ SYM is, see \eqref{(2.2.4)},\eqref{(2.3)},
\bq
\langle{\hat\Lambda}_{N_H}^{3}\rangle=\langle\Lambda_H^{2}\rangle m_H=\langle S\rangle=\la^2 \Bigl ( m_L^{N_L} m_H^{N_H} \Bigr )^{1/N_c}=\langle\lym^{3}\rangle\,,\quad {\hat\Lambda}_{N_H}^{3}=\Biggl (\frac{\la^{2N_F} m_H^{N_H}}{\det M_L}\Biggr )
^{1/N_H}\,,    \label{(2.2.7)}
\eq
\bbq
\Bigl (\frac{m^{\rm pole}_H\sim m_H}{\langle\Lambda_H^{2}\rangle} \Bigr ) \gg 1\,\ra\, r^{N_L/N_c}\ll \frac{m^2_H}{\la^2}\ll 1 \, \ra\,
\Bigl (\frac{\langle\lym^{3}\rangle}{m_H} \Bigr )=\Bigl ( \frac{\la}{m_H}\Bigr )^2\Biggl [ r^{N_L/N_c} \ll \frac{m^2_H}{\la^2} \Biggr ] \ll 1\,.
\eeq

Lowering the scale down to $\sim\langle\lym\rangle$ and integrating inclusively all $SU(N_H)$ gluons at $\mu=\langle \lym\rangle$ via the VY-procedure \cite{VY}, the lower energy Lagrangian at $\mu=\langle \lym\rangle$ looks as
\bq
K=2 z_Q(\la,\mu_{\rm gl,L}) {\rm Tr}\sqrt{(M^L_L)^\dagger M^L_L}+z_Q(\la,\mu_{\rm gl,L}){\rm Tr}\Bigl ( (\pi^H_L)^\dagger \pi^H_L+(\pi_H^L)^\dagger
\pi_H^L\Bigr )\,, \label{(2.2.8)}
\eq
\bbq
{\cal W}=\Biggl [ N_H ({\hat\Lambda}_{N_H})^3=N_H\Biggl (\frac{\la^{2N_F} m_H^{N_H}}{\det M_L}\Biggr )^{1/N_H}\Biggr ]+
m_L{\rm Tr} M_L + m_H {\rm Tr}\Bigl ( \pi^H_L \pi^L_H  \Bigr)\,, \quad \langle {\cal W}\rangle=N_c\langle S\rangle_{N_c}\,.
\eeq
\bbq
M^L_L=\langle M_L\rangle+\langle M_L\rangle^{1/2}\pi^L_L\,,\quad  2\,{\rm Tr}\sqrt{(M^L_L)^{\dagger} M^L_L}\ra {\rm Tr}\Bigl (\,(\pi^L_L)^\dagger \pi^L_L \Bigr )\,,
\eeq
where $\pi^L_L$  and $\pi^L_H,\, \pi^H_L$  fields are canonically normalized at $\mu=\la$, see \eqref{(2.2.3)}. At $\mu < \mu_{\rm gl,L}$ the RG-evolution of $N^2_L$ pions $M^L_L$ and $2 N_LN_H$ hybrid pions $\pi^H_L,\,\pi^L_H$ is frozen, see \eqref{(2.2.5)}.

From \eqref{(2.2.8)}, the particle masses are
\bq
\mu^{\rm pole}\Bigl (\pi^L_L ({\rm adj}) \Bigr )=\frac{m_L}{ z_Q(\la,\mu_{\rm gl,L})} \ll \langle\lym\rangle\,,\quad
\mu^{\rm pole}\Bigl (\pi^L_L ({\rm singl}) \Bigr )= \Bigl ( \frac{N_L}{N_H}+1\Bigr )\frac{m_L}{ z_Q(\la,\mu_{\rm gl,L})}\,, \label{(2.2.9)}
\eq
\bbq
M^L_L ({\rm singl})=\frac{1}{N_L}\sum_{i=1}^{N_L} M^i_i,\,\,\,\, \langle M^L_L ({\rm singl})\rangle=\la^2\Bigl ( \frac{1}{r}\Bigr )^{\frac{N_H}{N_c}}\,,\quad
\mu^{\rm pole} (\pi^L_H)=\mu^{\rm pole} (\pi_L^H)=\frac{m_H}{ z_Q(\la,\mu_{\rm gl,L})}=m^{\rm pole}_{H,hybr}\,.
\eeq

The mass spectrum looks in this case as follows. - \\
a)\, There are $N_L(2N_c - N_L)$\, ${\cal N}=1$ multiplets of real heavy gluons with masses $\mu_{\rm gl,L}\gg\la$, see \eqref{(2.2.2)}. $N_L^2$ of $(A_\mu)^L_L$ multiplets are weakly coupled and not confined, while  $2 N_L N_H$ heavy hybrids $(A_{\mu})^{L}_{H},\,\,(A_{\mu})_{L}^{H}$  multiplets are weakly coupled and weakly confined by the low energy $SU(N_H)$ SYM.\\
b)\,There are $2 N_L N_H$ $\,{\cal N}=1$ multiplets of complex genuine hybrid pseudo-Goldstone pions $\pi^L_H, \pi^H_L$ with masses $\mu^{\rm pole}(\pi^L_H)=\mu^{\rm pole}(\pi_L^H)=m^{\rm pole}_{H,hybr}$ which are not confined, see \eqref{(2.2.9)}.  \\
c)\, There is a number of hadrons made from weakly coupled and weakly confined by $SU(N_H)$ SYM active  quarks $\hh, \ohh$ with non-higgsed colors, with the typical mass scale ${\cal O} (m^{\rm pole}_H)\gg\langle\Lambda_H\rangle=\langle M\rangle^{1/2}_H$, see \eqref{(2.2.6)}. The genuine quark-antiquark bound states have masses $\approx 2 m^{\rm pole}_H$, and $N^2_H-1$\, ${\cal N}=1$ multiplets of genuine complex $\pi^H_H(adj)$ pseudo-Goldstone pions are among them.\\
 d)\,There is a number of gluonia of $SU(N_H)$ SYM with the typical mass scale ${\cal O} (\langle\lym\rangle)$, see \eqref{(2.2.7)}.\\
e)\, The lightest are $N_L^2\,\, \,{\cal N}=1$ multiplets of genuine complex pseudo-Goldstone pions $\pi^L_L(adj)$ with masses \eqref{(2.2.9)}.

Because the tension of the $SU(N_H)$ confining string is $\sigma^{1/2}\sim\langle\lym\rangle\ra 0$ \eqref{(2.2.7)}, {\it there is no confinement in the chiral limit $m_{L,H}\ra 0$ with fixed $r=m_L/m_H \ll 1$}.

\vspace*{3mm}

{\bf B)}\, If $m_H \ll \langle\Lambda_H\rangle\ll\la$, then active HH-quarks $\hh,\,\ohh$  with non-higgsed colors are also  in the HQ-phase, i.e. not higgsed but confined, but theory enters now at $\langle\lym\rangle < \mu < \langle\Lambda_H\rangle$ the perturbative strong coupling regime with the large gauge coupling $a(\langle\lym\rangle \ll \mu\ll\langle\Lambda_H\rangle)\gg 1$. These active  HH-quarks decouple then as heavy at  $\mu < m^{\rm pole}_H$, see section 3 in \cite{ch21}.

{\it The only difference with previous case $m_H\gg \Lambda_H$ is in the value of perturbative pole mass} $m^{\rm pole}_H$,
\bbq
m_H(m^{\rm pole}_H < \mu < \langle\Lambda_H\rangle)\sim \frac{m_H}{z_Q(\langle\Lambda_H\rangle, m^{\rm pole}_H < \mu < \langle\Lambda_H\rangle)},\,\, z_Q(\langle\Lambda_H\rangle, m^{\rm pole}_H < \mu < \langle\Lambda_H\rangle)\sim \Bigl (\frac{\mu}{\langle\Lambda_H\rangle}\Bigr )^{\gamma^{\rm str}_Q\, >\, 2} <  1,
\eeq
\bq
m^{\rm pole}_H\sim\frac{m_H}{z_Q(\langle\Lambda_H\rangle, m^{\rm pole}_H)}\,\ra \,
\langle\lym\rangle \ll m^{\rm pole}_H\sim \langle\Lambda_H\rangle\Bigl (\frac{m_H}{\langle\Lambda_H\rangle} \Bigr )^{0\, < \frac{1}{1+\gamma^{\rm str}_Q}\, < \frac{1}{3}\,}\ll \langle\Lambda_H\rangle\,.  \label{(2.2.10)}
\eq

After these active HH-quarks decoupled at $\mu=m^{\rm pole}_H$\, \eqref{(2.2.10)}, the scale factor ${\langle\hat\Lambda_{N_H}}\rangle$ of the $SU(N_H)$ SYM  gauge coupling is determined from matching, see \eqref{(2.1.2)}-\eqref{(2.1.3)},
\bq
a_{+}(\mu=m^{\rm pole}_H)=\Bigl (\frac{\langle\Lambda_{H}\rangle}{m^{\rm pole}_H} \Bigr )^{\nu=(\gamma^{\rm str}_Q-2)\,>\,0}=a^{(str,\, pert)}_{SYM}(\mu=m^{\rm pole}_H)=\Bigl (\frac{m^{\rm pole}_H)}{\langle\hat\Lambda_{N_H}\rangle} \Bigr )^3\, \ra  \label{(2.2.11)}
\eq
\bbq
\ra \,\,\langle{\hat\Lambda_{N_H}}^3\rangle\sim \la^2 r^{N_L/N_c}  m_H =\la^2 \Bigl ( m_L^{N_L} m_H^{N_H} \Bigr )^{1/N_c}=\langle\lym^3\rangle\,,
\eeq
as it should be, see \eqref{(2.3)}.

Besides,  as for a potentially possible higgsing of these HH-quarks, see \eqref{(2.1.1)}-\eqref{(2.1.4)},\eqref{(2.2.4)},\eqref{(2.2.10)}
\bq
K_{N_H}(\mu=\langle\Lambda_H\rangle)=Z_H {\rm Tr}_{N_H}\Bigl ( (\hh)^\dagger \hh + (\ohh)^\dagger \ohh \Bigr ),\quad
Z_H\sim \log^{-1/2} (\la/\langle\Lambda_H\rangle) \ll 1\,, \label{(2.2.12)}
\eq
\bbq
\frac{\mu^2_{{\rm gl},\,H}(m^{\rm pole}_H < \mu < \langle\Lambda_H\rangle)}{\mu^2}\sim \frac{Z_H}{\mu^2}\Biggl [ a(\langle\Lambda_H\rangle, \mu) z_Q(\langle\Lambda_H\rangle,\mu)\sim\frac{\mu^2}{\langle\Lambda^2_H\rangle}\Biggr ]\langle M_H\rangle \sim \frac{\mu^2_{{\rm gl},\,H}(\mu=m^{\, pole}_H)}{m^{\rm 2,\, pole}_H}\sim Z_H\ll 1\,.
\eeq

It is seen from \eqref{(2.2.12)} that, even if these HH-quarks were higgsed, they would be unable to give such mass to $SU(N_H)$ gluons which would stop the perturbative RG-evolution and there would be no pole in the gluon propagator at $m^{\rm pole}_H < \mu < \langle\Lambda_H\rangle$. And $\mu^2_{{\rm gl},H}(\mu=m^{\rm 2,\, pole}_H)/m^{\rm 2,\, pole}_H \ll 1$. This shows that these HH-quarks are really in the HQ (heavy quark) phase, i.e. not higgsed but confined by $SU(N_H)$ SYM. At
$\mu < m^{\rm pole}_H$ these active HH-quarks decouple.

And  as a check of self-consistency, see \eqref{(2.2.10)},
\bbq
\Bigl (\frac{m_H}{\langle\Lambda_H\rangle} \Bigr ) \ll 1\,\ra\, \Biggl [ \Bigl (\frac{1}{r}\Bigr )^{N_L/2N_c}\ll\frac{\la}{m_H}\,\Biggr ] \ra
\Bigl (\frac{{\langle\hat\Lambda_{N_H}\rangle}=\langle\lym\rangle}{m^{\rm pole}_H} \Bigr )^{3}\sim \Biggl [\frac{m_H}{\la} \Bigl (\frac{1}{r}\Bigr )^{N_L/2N_c}\ll 1 \Biggr ]^{\frac{(\gamma^{\rm str}_Q-2) =\, \nu}{(1+\gamma^{\rm str}_Q)}\, > \,\, 0}\ll 1,
\eeq
as it should be.

The mass spectrum in this case '{\bf B}' differs from '{\bf A}' only in the HH-sector which is strongly coupled now. Confined by $SU(N_H)$ SYM active quarks $\ohh,\,\hh$ are strongly coupled and weakly confined  and form a number of HH-hadrons with the typical mass scale ${\cal O}(m^{\rm pole}_H)$ \eqref{(2.2.10)}. And $N^2_H-1$\, ${\cal N}=1$ multiplets of genuine complex $\pi^H_H(adj)$ pseudo-Goldstone pions are among these HH-hadrons.

{\it The typical mass scale ${\cal O}(m^{\rm pole}_H)$\,  \eqref{(2.2.10)} of HH-hadrons is checked by \eqref{(2.2.11)}. Integrating inclusively all these HH-hadrons (i.e. equivalently, integrating inclusively all $\ohh,\,\hh$  quarks as heavy at $\mu= m^{\rm pole}_H$)\, \eqref{(2.2.10)}, we obtain the well known beforehand right value of $\langle\lym\rangle$}, see \eqref{(2.2.11)}.\\

In the chiral limit $m_{L,H}\ra 0$ with fixed $0 < r=m_L/m_H\ll 1$,\,\,   $\langle {\hat Q}^i_{\beta}(0)\rangle=\delta^i_{\beta}\,\rho_{\rm higgs}$ in \eqref{(2.2.2)} breaks {\it spontaneously} global $SU(N_F)_A\ra SU(N_H)_A$, so that there will be $N_L^2$ $\,\,{\cal N}=1$ multiplets of massless genuine complex Goldstone pions $\pi^L_L$ (see section 4) and $2N_L N_H$ $\,\,{\cal N}=1$ multiplets of massless genuine complex Goldstone hybrid pions $\pi^L_H+\pi^H_L$. (The supersymmetry doubles the number of real Goldstone particles).

All particle masses in the HH-sector also coalesce to zero, in spite of the strong gauge coupling. Because the tension of the $SU(N_H)$ confining string is $\sigma^{1/2}\sim\langle\lym\rangle\ra 0$ \eqref{(2.2.11)}, {\it there is no confinement} in this chiral limit, see footnotes \ref{(f1)},\,\ref{(f4)}.

For more details see section 4.

\numberwithin{equation}{section}

\section{Dual Seiberg's theory with $N_F=N_c$}

Let us recall first once more that in his paper \cite{S1} N. Seiberg proposed the $N_F=N_c$\, ${\cal N}=1$ SQCD  theory with $N^2_F$ mesons $M^i_j$ and two $B,\,{\ov B}$ baryons (with one constraint), and $N_F=N_c+1$ SQCD theory with $N_F^2=(N_c+1)^2$ mesons $M^i_j$ and $2 N_F=2 (N_c+1)$ baryons $B_i,\,\,{\ov B}^j$ as {\it the low energy forms at the scale $\mu < \la$ of, respectively, direct SQCD theories with  $N_F=N_c$ and  $N_F=N_c+1$ light quark flavors with $m_{Q,i}\ll\la$}. In \cite{IS} the regime at $N_F=N_c$ and $\mu < \la$ was called ''confinement with the (spontaneous, at
$m_{Q,i}\ra 0$) chiral symmetry breaking''  and those at $N_F=N_c+1$ as ''confinement without the chiral symmetry breaking''\,.

This implied that in these two strongly coupled at $\mu\sim\la$ direct theories the light direct quarks and gluons are confined by strings with the tension $\sigma^{1/2}\sim\la$ and form hadrons with masses $\sim\la$, and there remain at lower energies only corresponding light mesons and baryons. (These regimes were called latter in the literature as 'S-confinement'\,).\\

This proposal and interpretation was criticized in \cite{ch1} (see pages 18 and 19 in arXiv:0712.3167\, [hep-th]) and in \cite{ch3} (see Introduction and sections 7-9  in arXiv:0811.4283 [hep-th]). The reason was the following. The confinement originates {\it only} from SYM sector, see footnote \ref{(f1)}. And the $SU(N_c)$ SYM theory has only one dimensional parameter $\lym=(\la^{3 N_c-N_F} \Pi_{i=1}^{N_F}{m_{Q,i}})^{1/3 N_{c}}\ll\la$ at $m_{Q,i}\ll\la$. Therefore, it can not give the string with the tension $\sigma^{1/2}\sim\la$, but only with $\sigma^{1/2}\sim\lym\ll\la$. And e.g. direct light equal mass quarks and gluons remain light and {\it weakly confined by strings with $\sigma^{1/2}\sim\lym\ll\la$ and can't form hadrons with masses} $\sim\la$, see section 2.1\,.

For this reason, in our papers \cite{ch1,ch3} and in this paper we consider at $\mu < \la$ the direct theories with quarks and gluons and proposed by N. Seiberg \cite{S1} dual theories with mesons and baryons as {\it independent theories}. Then, the mass spectra of the direct and dual theories have to be calculated and compared, to see whether they are equivalent or not.

It was shown explicitly in Appendix 'B' in \cite{Session} how the proposed by N. Seiberg low energy dual theory with $N_F^2=(N_c+1)^2$ mesons $M^i_j$ and $2 N_F=2 (N_c+1)$ baryons $B_i,\,\,{\ov B}^j$ originates not as the low energy form of the direct $SU(N_c)$ theory with $N_F=N_c+1$, but from his dual $SU({\ov N}_c=N_F-N_c=2)$ SQCD theory \cite{S2} with $N_F=N_c+2$ dual quarks at the large mass of last direct quark $m_{Q,i=N_c+2}=\la$. Similarly, it is shown below how the proposed by N. Seiberg in \cite{S1} dual $N_F=N_c$ SQCD  theory with $N_c^2-1$ mesons $M^i_j$ and $B,\,{\ov B}$ baryons (but with different Kahler terms) originates not as the low energy form of the direct $SU(N_c)$ theory with $N_F=N_c$, but from the dual $N_F=N_c+1$ theory in \cite{Session} at the large mass of last direct quark $m_{Q,i=N_c+1}=\la$.

\numberwithin{equation}{subsection}

\subsection{Light equal mass quarks}

Let us remind now the properties of the Seiberg dual $\,{\cal N}=1$ SQCD theory with $N^2_F=(N_c+1)^2$ mesons $M^i_j$,\,\, $2 (N_c+1)$ baryons $B_i,\,\, {\ov B}^i$ and $\mi=m_Q \ll\la,\,\, i=1,...,N_c,\,\,m_{Q,N_c+1}=\la$, which was obtained from Seiberg's dual $SU({\ov N}_c=N_F-N_c=2)$ $\,\,N_F=N_c+2$ theory \cite{S2} increasing mass of one quark in the direct $SU(N_c)\,\, N_F=N_c+2$ theory, see appendix 'B' in \cite{Session},
and \eqref{(2.1)} above.

The dual Kahler terms at $\mu=\la$ look as
\bq
{\wt K}_{N_c+1}= {\rm Tr}_{N_c+1}\frac{M^\dagger M}{\la^2}+\sum_{i=1}^{N_F=N_c+1}\Bigl ( (B^\dagger)^i B_i +({\ov B}^\dagger)_i\,{\ov B}^{\,i}\Bigr ),
\label{(3.1.1)}
\eq
and the dual superpotential is
\bq
{\wt {\cal W}}_{N_c+1}=m_Q\sum_{i=1}^{N_c}  M^i_i +\la M^{N_c+1}_{N_c+1}+{\rm Tr}_{N_c+1}\,({\ov B}\frac{M}{\la} B)-\frac{\det_{N_c+1} M}{\la^{2N_c-1}}\,,\label{(3.1.2)}
\eq
where the baryons $B_i$ and ${\ov B}^i$ with $N_F=N_c+1$ flavors are really the remained light dual quarks, see appendix 'B' in \cite{Session}\,:
$B_i=\sum_{\alpha,\beta=1}^{2}\epsilon_{\alpha\beta}[\,\langle {\hat q}^{\,\alpha}_{j=N_c+2}\rangle/\la=\delta^{\alpha,1}]\,{\hat q}^{\,\beta}_i=
{\hat q}^{\,\beta=2}_i,\,\,i=1,...,N_c+1$ (and similarly ${\ov B}$).

After most dual particles with masses $\sim\la$ decoupled at $\mu < \la$, see \eqref{(3.1.2)}, there remained only $N_c^2+1$  mesons $M^i_j,\,\, i,j=1,...,N_F=N_c$ and $M^{N_c+1}_{N_c+1}$, and two baryons $B=B_{i=N_c+1}$ and ${\ov B}={\ov B}^{\,i=N_c+1}$. The lower energy Lagrangian looks as
\bq
{\wt K}_{N_c}(\mu=\la)= {\rm Tr}_{N_c}\frac{M^\dagger M}{\la^2}+ \frac{\Bigl ({M^{N_c+1}_{N_c+1}}\Bigr )^{\dagger}\,M^{N_c+1}_{N_c+1}}{\la^2}
+ (B^\dagger B+B\ra {\ov B})\,, \label{(3.1.3)}
\eq
\bq
{\wt {\cal W}}_{N_c}=M^{N_c+1}_{N_c+1}\Bigl (\la-\frac{\det_{N_c} M}{\la^{2N_c-1}} +\frac{{\ov B} B}{\la}\Bigr )+m_Q\sum_{i=1}^{N_F=N_c} M^i_i+...\,\, \label{(3.1.4)}
\eq
Dots in \eqref{(3.1.4)} denote terms which are irrelevant in what follows for our purposes.

The mean vacuum values  from \eqref{(3.1.3)}.\eqref{(3.1.4)} look as, compare with \eqref{(2.1)}\,,\eqref{(2.4)}\,,\eqref{(2.1.6)},
\bq
\frac{\langle\det_{N_c} M\rangle=\Pi_{i=1}^{N_F=N_c}\langle M^i_i\rangle}{\la^{2N_c-1}}=\la, \quad
\langle M^i_i\rangle=\la^2\,, \quad i=1,...,N_F\,, \quad\langle B\rangle=\langle{\ov B}\rangle=0\,,  \label{(3.1.5)}
\eq
\bq
M^i_j=M^i_j(adj)+ \delta^i_j\, \Bigl ( M(singl)\equiv M_s \Bigr ), \quad M_s=\frac{1}{N_c}\sum_{i=1}^{N_F=N_c} M^i_i\,, \quad
\langle M^i_j(adj)\rangle=0\,, \quad \langle M_s \rangle=\la^2\,,\label{(3.1.6)}
\eq
\bq
\langle \frac{\partial {\wt {\cal W}}_{N_c}}{\partial M^i_i}\rangle=0 \,\, \ra \,\, \langle M^{N_c+1}_{N_c+1}\rangle=m_Q\la=\frac{\langle S\rangle_{N_c}}{\la}\,,
\quad \langle {\wt {\cal W}}_{N_c}\rangle=m_Q\sum_{i=1}^{N_c}\langle M^i_i\rangle=N_c m_Q\la^2=N_c\langle S\rangle_{N_c}\,, \label{(3.1.7)}
\eq
see \eqref{(2.3)}. Because colorless $\langle B\rangle$ and $\langle{\ov B}\rangle$ depend analytically on $m_Q\neq 0$, they remain zero also in the limit $m_Q\ra 0$.

From \eqref{(3.1.3)},\eqref{(3.1.4)} the masses of heavier particles look as
\bq
\mu^{\rm pole}\Bigl ( M_s\Bigr ) = \mu^{\rm pole}\Bigl ( M^{N_c+1}_{N_c+1}\Bigr )\sim \la\,, \label{(3.1.8)}
\eq
while the masses of lighter particles look as (up to logarithmic renormalization factors)
\bq
 \mu^{\rm pole}\Bigl (M^i_j(\rm adj)\Bigr )\sim m_Q\ll \la,  \quad \mu^{\rm pole}(B)=\mu^{\rm pole}({\ov B})\sim m_Q\ll\la\,.\label{(3.1.9)}
\eq

Let us put attention that in \cite{S1} the Kahler terms of two physical fields $\delta M^{N_c+1}_{N_c+1}=(M^{N_c+1}_{N_c+1}-
\langle M^{N_c+1}_{N_c+1}\rangle\,)$ and $\delta M_s=(M_s-\langle M_s\rangle\,)$ are excluded by hands from \eqref{(3.1.3)} (in order to match the 't Hooft triangles). Then $\delta M^{N_c+1}_{N_c+1}$ becomes the auxiliary field and the physical field $\delta M_s$ is excluded by the constraint $\det_
{N_c} M-\la^{2N_c-2}\,{\ov B}B   =\la^{2N_c}$.  While here these two fields are normal physical fields and this looks at least much more natural. \\

In the chiral limit $m_Q\ra 0$ the dual Lagrangian \eqref{(3.1.3)},\eqref{(3.1.4)} is $SU(N_F=N_c)_{left}\times SU(N_F=N_c)_{right}\times U(1)_{B,V}\times U(1)_R$ invariant. (The R-charges of scalars $M$ and $B,\,{\ov B}$ in \eqref{(3.1.4)} are zero, R-charge of scalar $M^{N_c+1}_{N_c+1}$ is 2). But $\langle M_s\rangle=\la^2$ \eqref{(3.1.6)} breaks spontaneously $SU(N_F=N_c)_A$. And $N_F^2-1$ ${\cal N}=1$ multiplets of Goldstone particles $M^i_j(\rm adj)$ become massless in the chiral limit.

The symmetry $U(1)_{N_F=N_c,A}$ is broken explicitly by the term $\sim \det_{N_F=N_c}M$ in \eqref{(3.1.4)}. {\it The coupling  of $M_s$ with $M^{N_c+1}_{N_c+1}$ gives them both large masses} $\sim\la$.

 In this limit $m_Q\ra 0$, the whole moduli 'space' is exhausted by one point:\, $\langle M_s\rangle=\la^2,\,\, \langle M^i_j(\rm adj)\rangle=0, \,\,
\langle B\rangle=\langle{\ov B}\rangle=0$.

And finally, about the  anomalous 't Hooft triangles. In the chiral limit $m_Q\ra 0$ all particle masses of the direct theory coalesce to zero. In the dual theory  masses of $\mu^{\rm pole}\Bigl (M^i_j(\rm adj)\Bigr )$ and $\mu^{\rm pole}(B)=\mu^{\rm pole}({\ov B})$ tend to zero \eqref{(3.1.9)}, while
$\mu^{\rm pole}\Bigl ( M_s\Bigr ) = \mu^{\rm pole}\Bigl ( M^{N_c+1}_{N_c+1})\Bigr )\sim \la$ \eqref{(3.1.8)}. It is not difficult to check that at scales $\mu\ll\la$  all 't Hooft triangles of direct and dual theories are matched, see \cite{Amati},\cite{S1}.

At $0 < m_Q\ll\la $ the 't Hooft triangles are matched only in the range of scales $m^{\rm pole}_Q\ll\mu\ll\la$ \eqref{(2.1.1)}, where all particles of the direct and dual theories are effectively massless. At scales $\mu < m^{\rm pole}_Q$ triangles are not matched because the mass spectra of direct and dual theories are qualitatively different.

\numberwithin{equation}{subsection}
\subsection{Light unequal mass quarks}

Consider now the case with unequal masses\,: $\mi=\ml,\,\,i=1,...,N_L,\,\,\,m_{Q,k}=\mh,\,\, k=N_L+1,...,N_c,\\ N_L+N_H=N_F=N_c$,
$0 < m_L\ll m_H\ll\la$,  with fixed $0 < r=m_L/m_H\ll 1$.

The Lagrangian at $\mu=\la$ is taken as
\bq
{\wt K}_{N_c+1}= {\rm Tr}_{N_c+1}\frac{M^\dagger M}{\la^2}+\sum_{i=1}^{N_F=N_c+1}\Bigl ( (B^\dagger)^i B_i +({\ov B}^\dagger)_i\,{\ov B}^{\,i}\Bigr ),
\label{(3.2.1)}
\eq
\bbq
{\wt {\cal W}}_{N_c+1}=m_L\sum_{i=1}^{N_L}  (M_L)^i_i + m_H\sum_{i=1}^{N_H}  (M_H)^i_i + \la M^{N_c+1}_{N_c+1}+{\rm Tr}_{N_c+1}\,({\ov B}\frac{M}{\la} B)-\frac{\det_{N_c+1} M}{\la^{2N_c-1}}\,.
\eeq

From \eqref{(3.2.1)}\,:
\bq
\langle M_L\rangle=\la^2\Bigl (\frac{1}{r} \Bigr )^{N_H/N_c}\gg\la^2\,,\quad \langle M_H\rangle=\la^2\Bigl (r \Bigr )^{N_L/N_c}\ll \la^2\,, \label{(3.2.2)}
\eq
\bbq
\langle (\det M)_{N_c} \rangle=\langle M_L\rangle^{N_L}\langle M_H\rangle^{N_H}=\la^{2N_c}\,,\quad \langle M^{N_c+1}_{N_c+1}\rangle=\frac{m_L\langle M_L\rangle}{\la}=\frac{m_H\langle M_H\rangle}{\la}=\frac{\langle S\rangle_{N_c}}{\la}\,,
\eeq
\bbq
\langle (\det M)_{N_c+1} \rangle=\la^{2N_c}\Biggl [\langle M^{N_c+1}_{N_c+1}\rangle=\frac{\langle S\rangle_{N_c}}{\la}\Biggr ]\,,\quad 0 < r=m_L/m_H \ll 1\,,
\eeq
compare with \eqref{(2.2.1)},
\bq
\langle B_L\rangle=\langle{\ov B}_L\rangle=\langle B_H\rangle=\langle{\ov B}_H\rangle=\langle B\rangle=\langle{\ov B}\rangle=0\,,  \label{(3.2.3)}
\eq
compare with \eqref{(3.1.5)},
\bq
\langle {\wt {\cal W}}\rangle=m_L\sum_{i=1}^{N_L}\langle M_L\rangle + m_H\sum_{i=1}^{N_H}\langle M_H\rangle=N_c\langle S\rangle_{N_c}\,, \quad
\langle S\rangle_{N_c}=\la^2 m_L^{N_L/N_c} m_H^{N_H/N_c}=r^{N_l/N_c}\la^2 m_H \,, \label{(3.2.4)}
\eq
compare with \eqref{(2.1.6)}, \eqref{(2.2.8)},\eqref{(3.1.7)}.

The particle masses look at $\mu\sim\la$ as, see \eqref{(3.2.1)},\eqref{(3.2.2)},
\bq
\mu(B_L)=\mu({\ov B}_L)\sim\frac{\langle M_L\rangle}{\la}\sim \la\Bigl (\frac{1}{r}\Bigr )^{N_H/N_c}\gg\la\,,\quad \mu(B_H)=\mu({\ov B}_H)
\sim\frac{\langle M_H\rangle}{\la}\sim\la\Bigl (r \Bigr )^{N_L/N_c}\ll \la,\,\,\,  \label{(3.2.5)}
\eq
\bq
\quad \mu(B=B_{i=N_c+1})=\mu({\ov B}={\ov B}^{i=N_c+1})\sim \frac{\langle S\rangle_{N_c}}{\la^2}\sim r^{N_L/N_c} m_H \ll\la\,, \label{(3.2.6)}
\eq
\bbq
(M^L_L)^i_j=(M^L_L)(\rm adj)^i_j+ \delta^i_j (M_s)^L_L,\,\, (M_s)^L_L =\frac{1}{N_L}\sum_{i=1}^{N_L} M_{L,i}, \,\,
(M^H_H)^i_j=(M^H_H(\rm adj)^i_j+ \delta^i_j (M_s)^H_H,\,\, (M_s)^H_H=\frac{1}{N_H}\sum_{i=1}^{N_H} M_{H,i},
\eeq
\bbq
\langle M^i_j(\rm adj)\rangle=0\,, \quad \langle (M_s)^L_L \rangle=\langle M_L\rangle\,, \quad \langle (M_s)^H_H \rangle=\langle M_H\rangle\,,
\eeq
\bq
\mu\Bigl ( (M_s )^H_H\Bigr )\sim \mu\Bigl (M^{N_c+1}_{N_c+1} \Bigr )\sim\frac{\la}{r^{N_L/N_c}}\gg\la\,,
\quad \mu\Bigl ( (M_s )^L_L\Bigr )\sim  \,m_L \ll\la\,, \label{(3.2.7)}
\eq
\bq
\mu\Bigl (M^L_L(\rm adj)\Bigr )\sim \frac{\langle (M^{N_c+1}_{N_c+1} \rangle \la^3}{\langle M^L_L\rangle^2}\sim
 r^{(N_c+N_H)/N_c}\,m_H\ll\la\,,\quad  \mu\Bigl (M^H_H(\rm adj)\Bigr )\sim \frac{m_H}{ r^{ N_L/N_c}}\lessgtr \la\,,\label{(3.2.8)}
\eq
\bq
\mu\Bigl ( M^L_H(\rm adj)\Bigr )=\mu\Bigl ( M_L^H(\rm adj)\Bigr )\sim  r^{N_H/N_c} m_H\ll\la\,. \label{(3.2.9)}
\eq
Therefore, the heavy particles with masses $\gg\la$ at small $\mi\neq 0$ and $r\ll 1$ are: $B_L,\,{\ov B}_L,\, (M_s)^H_H, \, M^{N_c+1}_{N_c+1}$ and maybe $M^H_H(\rm adj)$, while the particle with masses $\ll\la$ are
\bq
B_H,\,\,{\ov B}_H,\,\, B,\,\, {\ov B},\,\,(M_s)^L_L,\,\, M^L_L(\rm adj),\,\,M^L_H(\rm adj)=M^H_L(\rm adj),\,\,
{\rm and\,\, may\,\, be}\,\, M^H_H(\rm adj)\,. \label{(3.2.10)}
\eq

In the chiral limit $m_{L,H}\ra 0$ with fixed $0 < r=m_L/m_H\ll 1$, the Lagrangian ${\wt {\cal W}}_{N_c+1}$ \eqref{(3.2.1)} is invariant under global
$SU(N_F=N_c)_{left}\times SU(N_F=N_c)_{right}\times U(1)_{V,B}\times U(1)_R$. The mean values $\langle (M_s)^L_L\rangle =\langle M_L\rangle\neq \langle (M_s)^H_H\rangle=\langle M_H\rangle$  \eqref{(3.2.2)},\eqref{(3.2.6)} break spontaneously $SU(N_F=N_c)_{left}$ and  $SU(N_F=N_c)_{right}$. The residual symmetry is $SU(N_L)_V\times  SU(N_H)_V\times U(1)_{V,L}\times U(1)_{V,H}\times U(1)_R$. Therefore,  $N_L^2-1\,\,  M^L_L(\rm adj),\,\,N_H^2-1\,\,  M^H_H(\rm adj)$ and $2N_L N_H$ hybrids  are the genuine Goldstone particles and are massless in this limit, see \eqref{(3.2.8)}, \eqref{(3.2.9)}. Because the supersymmetry doubles the number of Goldstones, the spontaneous breaking of $SU(N_F=N_c)_V$ does not increase the number of Goldstones. The term  $\sim \det_{N_c}M$ in \eqref{(3.2.1)} is not invariant under $U(1)_{A,L}$ and $U(1)_{A,H}$.  The mass $\mu^{\rm pole}\Bigl ( (M_s )^L_L\Bigr )$ tends to zero in this limit, see \eqref{(3.2.7)}. But the mass of $(M_s)^H_H$ is nonzero in this chiral limit, see \eqref{(3.2.7)}. This differs from the direct theory, see section 2.2, where masses of the whole HH-sector tend to zero at $m_H\ra 0$.

The whole moduli space is exhausted in this limit by the line of different values of $0 < r < \infty$.

The 't Hooft triangles of the direct and dual theories are matched in this chiral limit at scales $\mu < \mu^{\rm pole }(B_H)$. But at $\mu^{\rm pole }(B_H) < \mu < \la$ \eqref{(3.2.5)} the baryons $B_H,\,{\ov B}_H$ give additional contributions to the triangles $R^1=R^3$ and $RB^2$ and spoil the matching.

At small $0 < m_L\ll m_H\ll\la$, the values of 't Hooft triangles depend on ranges of the scale considered. Because the mass spectra of particles with masses $\ll\la$ are qualitatively different in the direct and dual theories, the triangles are not matched.

\numberwithin{equation}{section}
\section{Direct theory:\, (pseudo)Goldstone particles}

{\bf A)} Equal quark masses. In the chiral limit $m_Q\ra 0$, the non-anomalous global symmetry of the Lagrangian \eqref{(2.2)} is $SU(N_c)\times SU(N_F=N_c)_{left}\times SU(N_F=N_c)_{right}\times U(1)_R\times U(1)_{V,B}$. Recall that {\it all quarks are not higgsed} in this case, see \eqref{(2.7)}. All quark masses $m^{\rm pole}_Q$  \eqref{(2.1.1)} tend to zero. Besides, $\langle M^i_j(adj)\rangle=0, \,\, \langle M({\rm singl})\rangle=\la^2$, see the end of section 2.1.

$\langle M({\rm singl})\rangle=\la^2$ breaks {\it spontaneously} the whole $SU(N_F)_A$ global symmetry, while the global $SU(N_F)_V$ and $U(1)_R\times U(1)_{V,B}$ symmetries remain unbroken (remind that R-charge of scalars $M^i_j$ is zero). As a result, there will be $N^2_F-1$\,\,${\cal N}=1$ complex multiplets of genuine (i.e. non-anomalous) massless Goldstone pions $\pi^i_j(adj)$ (the supersymmetry doubles the minimal number of real Goldstone fields predicted by the Goldstone theorem).

The global Abelian symmetry  $U(1)_{N_F,A}$ is also broken spontaneously by $\langle M({\rm singl})\rangle=\la^2$.  But this symmetry is anomalous, i.e. it is explicitly broken in addition.

As it is seen from \eqref{(2.3)}, the tension of the confining string $\sigma^{1/2}\sim \lym\ra 0$, i.e. {\it there is no confinement in this chiral limit} $m_Q\ra 0$. Moreover, the masses of all quarks and gluons ($\mu^{\rm non-pert}_{\rm gl}\sim\lym\ra 0$) and hadrons made from them coalesce to zero. I.e., {\it the whole mass spectrum consists only from massless quarks and gluons}, see footnotes \ref{(f1)},\,\ref{(f4)}. For these reasons,
all 't Hooft triangles are matched automatically.\\

At small $0 < m_Q\ll\la$ all quarks are still not higgsed but are weakly confined now by $SU(N_c)$ SYM, i.e. $\sigma^{1/2}\sim \lym\ll m^{\rm pole}_Q\ll\la$, see \eqref{(2.7)},\eqref{(2.1.1)},\eqref{(2.1.5)}. Hadrons made from them have typical masses ${\cal O} (m^{\rm pole}_Q)\ll\la$. And $N_F^2-1$ genuine pseudo-Goldstone pions $\pi^i_j(adj)$ are among these hadrons. All quarks and gluons on the one hand, and hadrons on the other one form two full bases in the same space. And, in the cases of {\it inclusive} integrations, we can use those basis which is more convenient. At scales ${\cal O} (m^{\rm pole}_Q)\ll \mu \ll\la$ all particles in the mass spectrum, i.e. hadrons, quarks and gluons, are effectively massless. And in this range of scales two sets of 't Hooft triangles, those at scales $\mu\gg\la$ and those at ${\cal O} (m^{\rm pole}_Q)\ll \mu \ll\la$,  are matched automatically in the basis of quarks and gluons.

At scales $\mu \ll m^{\rm pole}_Q$ 't Hooft triangles are changed because all quarks decouple as heavy, but this is normal. This is a result of not spontaneous but the {\it explicit}  $SU(N_F)_A$ global symmetry breaking.\\

{\bf B)} Unequal quark masses. When all quark masses tend to zero with fixed ratio $0 < r=m_L/m_H\ll1$ as in section 2.2, the global symmetry of the Lagrangian is the same as in {\bf A} above. But the spontaneous breaking is different, see \eqref{(2.2.1)}. {\it LL-quarks with e.g. $0 < N_L < N_c-1$
are now higgsed in the weak coupling regime}, $\langle {\hat Q}^i_{\beta}\rangle=\delta^i_{\beta} \rho_{\rm higgs},\,\,\, \langle {\hat {\ov Q}}_i^{\,\beta}\rangle=\delta_i^{\beta} \rho_{\rm higgs},\,\, i,\beta=1...N_L,\,\,\rho_{\rm higgs}\gg\la$, see \eqref{(2.2.1)},\eqref{(2.2.2)} and footnote \ref{(f5)}. Now, not only the whole global $SU(N_F)_{A,F}$ is broken spontaneously by higgsed LL-quarks and  $\langle (M_{L,H})({\rm singl})\rangle=\langle M_{L,H}\rangle\neq 0,\, \langle M_{L}\rangle\neq \langle M_{H}\rangle$\,, but also global $SU(N_c)_C\times SU(N_F)_{V,F}\times U(1)_{V,B}$. The unbroken non-anomalous part looks as, see \eqref{(2.2.1)}\,~:\, $SU(N_L)_{V,C+F}\times SU(N_H)_{V,F}\times U(1)_{V, B_H}\times U(1)_{{\tilde V}, F+C}\times U(1)_R$. Both $U(1)_{A,L}$ and (\,$U(1)_{A,H}$ are broken spontaneously but, besides, they both are anomalous.

Now, the vector baryon charge is redefined: $B\ra B_H$, such that $B_L$-charge is zero for all $2 N^2_L$  higgsed \, LL-quarks
$Q^i_{\beta},\, {\ov Q}^{\beta}_i,\,\,i,{\beta}=1...N_L$ with L-flavors and L-colors. And the nonzero generator of $B_H$-charge is normalized in the
same way as $B$\,: ${\rm Tr}\,(B^2)={\rm Tr}\, (B^2_H)$. I.e., the generators of vector $B,\, {\tilde V}$ and $B_H$ look as
\bq
B= diag\, (\,\underbrace{\,  1}_{N_F})\,,\quad {\tilde V}= diag\, (\,\underbrace{\, - 1}_{N_L}\,;  \,\underbrace{\, N_L/N_H}_{N_H}\,),\,\quad
B_H=(N_F/N_H)^{1/2}\, diag\, (\,\underbrace{\, 0}_{N_L}\,;  \,\underbrace{\, 1}_{N_H}\,)\,. \label{(4.1)}
\eq

All $2 N_L N_H$ hybrid $LH+HL$ generators of $SU(N_F)_{V,F}$ are broken spontaneously by $\langle (M_{L,H})({\rm singl})\rangle=\langle M_{L,H}\rangle\neq 0,\,\, \langle M_L\rangle\neq \langle M_H\rangle$ \,\eqref{(2.2.1)}. But the Goldstone theorem determines only a {\it minimal} number of genuine real Goldstone fields. While supersymmetry doubles the number of Goldstone fields. And so, spontaneously broken vector hybrid generators do not increase the number of genuine Goldstone fields.

Except for $N_L(2N_c-N_L)$\, ${\cal N}=1$ multiplets of heavy gluons, the masses of all other lighter particles, quarks, gluons and hadrons coalesce to zero in this chiral limit. And masses of $(N_F^2-1)$ ${\cal N}=1$ multiplets of genuine complex Goldstone fields among them. {\it Because $\langle\lym\rangle\ra 0$, there is no confinement}.
\footnote{\,
The massless HH genuine Goldstone particles, see sections 2.2, considered as bound state fields $\sum_{\beta=1}^{N_H}(\ohh)^{\beta}_j (\hh)^i_{\beta})$ of not higgsed massless HH quarks with approaching zero binding energy, have binding radius $R_{\rm bind}\ra \infty$ and are indistinguishable from the state of two unbound HH quarks. And this concerns the approaching to zero masses of all other HH hadrons made from massless HH quarks. \label{(f4)}
}

There are at small $0 < m_L\ll m_H\ll\la$ the following light fields (see section 2.2):\, a)\, $2 N_H^2$ complex ${\cal N}=1$ multiplets of  weakly confined  active $\hh,\,\ohh$ quarks with $N_H\times N_H$ color and flavor degrees of freedom. They form HH hadrons with typical masses ${\cal O}(m^{\rm pole}_H)$\,\, \eqref{(2.2.6)},\eqref{(2.2.10)}, and $N^2_H-1$\, ${\cal N}=1$ multiplets of genuine complex pseudo-Goldstone pions $\pi^H_H(adj)$ are among these hadrons;\,\,  b) ${\cal N}=1$ multiplet of $SU(N_H)$ gluons with non-perturbative masses $\sim\langle\lym\rangle$.\, Besides, there are not confined:\,\, c)\, $N_L^2$\, ${\cal N}=1$ multiplets of genuine complex $\pi^L_L(adj)$ pseudo-Goldstone fields with masses $\sim m_L\ll\la$ (with logarithmic accuracy, see \eqref{(2.2.9)}\,);\, d)\, $2 N_L N_H$ \, ${\cal N}=1$ multiplets of genuine complex pseudo-Goldstone hybrids $\pi^H_L,\,\pi^L_H$ with masses $\sim m_H\ll\la$ (with logarithmic accuracy, see \eqref{(2.2.9)}\,).

As for the 't Hooft triangles, see the end of section 3.2.

\numberwithin{equation}{section}
\section{About regimes on the moduli space}

Consider now the direct theory with $N_{L,H},\,\,N_L+N_H=N_F=N_c$ quarks with masses ${m_{L,H}\ra 0}$ at fixed ratio $0 < r=m_L/m_H =
{\rm const} < \infty$, see section 2,
\bq
{\cal W}_{\rm matter}=\sum_{i=1}^{N_c}\mi M^i_i \ra 0,\,\, \langle S\rangle=\langle\lym^3\rangle \ra 0,\,\,
 \langle \det M^i_j\rangle=\la^{2 N_c},\,\, \langle B\rangle=\langle {\ov B}\rangle=0,\,\, m_{L,H}\ra 0\,\,.\label{(5.1)}
\eq
The moduli $\langle M_{L,H}\rangle$ are fixed in this limit, see \eqref{(2.2.1)}.

The theory \eqref{(2.2)} approaches now, at least, to the subspace with $\langle B\rangle=\langle {\ov B}\rangle=0$ of the whole moduli space. And this will be sufficient for our purposes.

{\bf{A)}}\, The region $0 < r=m_L/m_H\ll 1$. The direct theory approaches in the limit $m_{L,H}\ra 0$ to those region of the moduli space where LL-quarks $Q^L_L, {\ov Q}^L_L$ are higgsed, see \eqref{(2.2.1)},\eqref{(2.2.2)},
\footnote{\,
Here and below: at not too large values of $N_c$, see \cite{ch21}. If, at fixed $0 < r\ll 1$,  $N_c$ is so large that $\mu_{\rm {gl,L}}/\la^2\ll 1$, see \eqref{(2.2.2)}, then even LL-quarks are in the HQ (heavy quark) phase, i.e. not higgsed but confined at $\mi\neq 0$. Then all quarks are in the HQ phase with $m^{\rm pole}_H\gg m^{\rm pole}_L\gg\lym$,\, $\lym$ is as in \eqref{(2.3)}, and with $\langle {\hat Q}^i_a\rangle=\langle {\hat{\ov Q}}_i^{\,a}\rangle=0$. And similarly for H-quarks at $r\gg 1$, see \eqref{(2.2.1)}. \label{(f5)}
}
while HH-quarks $Q^H_H,\, {\ov Q}^H_H$ with H-flavors and H-colors are not. And $m^{\rm pole}_H\ra 0$, see \eqref{(2.2.6)},\eqref{(2.2.10)}.  Moreover, the scale factor of the  unbroken by higgsed LL-quarks low energy non-Abelian $SU(N_c-N_L=N_H\geq 2)$ SYM gauge coupling $\langle{\hat\Lambda}_{N_H}\rangle=\langle\lym\rangle\, \ra\,0$, see \eqref{(2.2.7)},\eqref{(2.3)}.  Because the  confining $SU(N_H\geq 2)$ SYM theory has only one dimensional parameter $\langle\lym\rangle$, the tension of its confining string is $\sigma^{1/2}\sim \langle\lym\rangle \ra 0$. I.e., {\it there is no confinement in the direct theory in this region of the moduli space}.

{\bf{B)}}\, Let us now consider another region of the moduli space where $r=m_L/m_H\gg 1$. There, vice versa, see \eqref{(2.2.1)} and footnote \ref{(f5)}, quarks $Q^H_H,\, {\ov Q}^H_H$ are higgsed while quarks $Q^L_L,\, {\ov Q}^L_L$  with L-flavors and L-colors are not, with $N_L \leftrightarrow N_H,\,\,\mu^2_{\rm gl,\, L}\leftrightarrow \mu^2_{\rm gl,\,\, H},\,\, m^{\rm pole}_H\leftrightarrow m^{\rm pole}_L$. And also the scale factor of the  non-Abelian $SU(N_L\geq 2)$ SYM gauge coupling is $\langle\lym\rangle\, \ra\,0$ \eqref{(2.3)}.  I.e., {\it there is no confinement in the direct theory also in this another region of the moduli space}.\\

In the paper \cite{FS} of E. Fradkin and S.H. Shenker, the special (not supersymmetric) QCD-type lattice $SU(N_c)$ gauge theory with $N_F=N_c$ flavors of scalar quarks $\Phi^i_{\beta}$ in the bi-fundamental representation was considered. In the unitary gauge, all remained $N_c^2+1$ physical real degrees of freedom of these quarks were {\it deleted  by hands} and replaced by one constant parameter $|v| > 0\,:\,  \Phi^{i}_{\beta} = \delta ^{i}_{\beta}|v|,\,\, \beta=1,...,N_c,\,\, i=1,...,N_F=N_c$.\, I.e., all such "quarks" are massless, with no self-interactions  and {\it permanently higgsed by hands} even at small $g|v|\ll\Lambda_{QCD}$, see page 3694 and eq.(4.1) for the bare perturbative Lagrangian in \cite{FS}. (Here  
$\Lambda_{QCD}={\rm const}$ is the analog of $\la={\rm const}$ in ${\cal N}=1$ SQCD). And all $N^2_c-1$ electric gluons received  {\it fixed masses} $g |v|$. The region with the large values of $g|v|\gg\Lambda_{QCD}$ was considered in \cite{FS} as the higgs regime, while those with small $0 < g|v|\ll\Lambda_{QCD}$ as the confinement one. The conclusion of \cite{FS} was that the transition between the higgs and confinement regimes is the analytic crossover, not the non-analytic phase transition. And although the theory considered in \cite{FS} was very specific, the experience shows that up to now there is a widely spread  opinion that this conclusion has general applicability both to lattice and continuum theories, and to non-supersymmetric and supersymmetric ones. And not only for the QCD with "defective" quarks from \cite{FS}, {\it but also for normal scalar quarks with all their degrees of freedom}.\\

Let us note that this model of E. Fradkin and S.H. Shenker \cite{FS} with such {\it permanently higgsed by hands at all $\,0 < g |v| < \infty$ non-dynamical scalar "quarks" $\Phi^i_{\beta} = \delta ^i_{\beta}|v|$ looks unphysical and is incompatible  with  normal model with dynamical electrically charged scalar quarks 
$\phi^i_{\beta}$ with all $2 N_F N_c$ their real physical degrees of freedom}. This model \cite{FS} is really the Stueckelberg pure $SU(N_c)$ YM-theory {\it with no dynamical electric quarks and with  massive all $N_c^2-1$ electric gluons with fixed by hands masses} $g|v| > 0$ , see eq.(4.1) in \cite{FS}.

For this reason, {\it in any case}, the electric flux emanating from the test (anti)quark becomes {\it exponentially suppressed} at distances $L > l_0=(g|v|)^{-1}$ from the source. And so, {\it the potentially possible confining string tension will be also exponentially suppressed at distances $L > l_0$ from sources}. And e.g. external heavy test quark-antiquark pair will be not connected then by one common {\it really confining string} at large distance between them. These quark and antiquark can move then practically independently of each other and can be registered alone in two different detectors at large distance between one another. I.e., {\it in any case}, in this Stuckelberg theory \cite{FS}, at all fixed $|v| > 0$, {\it there is no genuine confinement} which prevents appearance of one (anti)quark in the far detector. See also page 9 in v6 of \cite{ch21}.\\

Now, about non-perturbative effects in the {\it standard non-supersymmetric} $SU(N_c)$ pure YM-theory. The common opinion (supported by lattice calculation) is that there are $N_c-1$ independent magnetic monopoles $M_n$ {\it condensing in the vacuum state with the density $\rho^2_{\rm magn}\sim\Lambda^2_{YM}$}.  Due to this condensation,  corresponding dual {\it magnetic photons acquire masses} $\mu_{\rm magn}\sim {\tilde g}\rho_{\rm magn}\sim{\tilde g}\Lambda_{\rm YM}$. And this leads to real confinement of normal electrically charged particles by the confining strings with the typical tension $\sigma^{1/2}\sim\Lambda_{\rm YM}$.

But, as was emphasized in \cite{ch21}, responsible for confinement condensed {\it magnetic  monopoles and mutually non-local with them electric scalar quarks can not condense simultaneously in the vacuum state}. For this reason, because the condensate of all defective electric "quarks" $\Phi^i_{\beta}=\delta ^i_{\beta}|v|$ is {\it strictly fixed by hands} at $\rho_{\rm electr}=|v| > 0$ in \cite{FS}, {\it this prevents mutually nonlocal with them magnetic monopoles to condense simultaneously with such
"quarks"\, in the vacuum state. I.e., definitely, there is then no confinement of electric charges at all $0 < |v| < \infty$ in the Stuckelberg $SU(N_c)$ YM-theory used in \cite{FS}. All electric charges are really not confined but screened}. Therefore, the conclusion of \cite{FS} about an analytical crossover between the "regime with confinement" at $0 < g|v|\ll\Lambda_{YM}$ (according \cite{FS}) and the higgs regime at $g|v|\gg\Lambda_{YM}$ is not surprising. {\it This Stuckelberg theory used in \cite{FS} with such defective non-dynamical electric "quarks" $\Phi^i_{\beta} = \delta ^i_{\beta}|v|,\,\, |v| > 0$ stays permanently in the completely higgsed by hands phase with massive all electric gluons, with no condensation of magnetic monopoles and no confinement of electrically charged particles}.\\

To describe all this in more details, we will need more detailed picture of the confinement mechanist of electric charges in the standard pure $SU(N_c)_c$  YM theory. We describe now in short the proposed model for this confinement  mechanism.
~\footnote{\,
Nevertheless, we expect that this model well may be realistic. \label{(f6)}
}

This mechanism is qualitatively similar to those which is realized in ${\cal N}=2$ SYM theory softly broken to ${\cal N}=1$ SYM by the small but nonzero  mass term $\mu_X {\rm Tr}\, (X^{\rm adj})^2$ of the $SU(N_c)_c$ adjoint scalar superfield $(X^{\rm adj})^i_j$

\cite{SW1},\cite{SW2},\cite{DS}. But, unlike the ${\cal N}=2$ SYM, where there is this {\it elementary}  adjoint field $(X^{\rm adj})^i_j$ which condenses in the vacuum resulting in $SU(N_c)_c\ra U(1)^{N_c-1}$, in the ordinary pure $SU(N_c > 2)_c$ YM this role is played by the {\it composite} adjoint field $(H^{\rm adj})^i_j(x)$ (see section 2 in \cite{ch21} for the meaning of the hat)
\bq
(H^{\rm adj})^i_j(x)=\frac{\Lambda_{YM}^{-3}}{16\pi}\sum_{A,B,C=1}^{N_c^2-1} d^{ABC}(T^{A})^i_j G^B_{\mu\nu}(x) G^{C,\,\mu\nu}(x)\,,\label{(5.2)}
\eq
\bbq
\langle ({\hat H}^{\rm adj})^i_j(0)\rangle= {\rm diag}\, (\rho_1,...,\rho_{N_c})\,, \quad \sum_{k=1}^{N_c} \rho_k=0\,, \quad  \rho_k=c_k(N_c)\Lambda_{YM},\quad c_k(N_c)\lesssim 1,
\eeq
where dimensionless coefficients $c_k(N_c)$ depend on $N_c$ while $\langle ({\hat H}^{\rm adj})^i_j(0)\rangle$ is the {\it colored gauge invariant order parameter}, see section 2 in \cite{ch21}. As a result, $SU(N_c)_c\ra U(1)^{N_c-1}$, all charged electric gluons acquire masses $\mu_{\rm electr,ij}= g(\rho_i-\rho_j)$  and there appear $N_c-1$ independent scalar magnetic monopoles $M_n$. {\it These monopoles also condense in the vacuum}  with the density $\rho^2_{{\rm magn},n}= d^{\,2}_n(N_c)\Lambda^2_{YM}$ and give masses $\mu_{\rm magn,n}=  {\tilde g}d_n(N_c) \Lambda_{YM}$ to all $N_c-1$ dual magnetic photons. This leads to the genuine confinement of all $N^2_c-N_c$ charged electric massive  $SU(N_c)_c$ gluons by strings with the tension $\sigma^{1/2}\sim \Lambda_{YM}$.
\footnote{\,
Instead of $(H^{\rm adj})^i_{j}(x)$ in \eqref{(5.2)}, in ${\cal N}=1\,\,SU(N_c > 2)$ SYM this will be its analog -  the chiral superfield  \\
$\hspace*{8mm} (h^{\rm adj})^i_{j}(x)=\Lambda_{SYM}^{-2}\sum_{A,B,C=1}^{N_c^2-1}\sum_{\gamma=1}^{2}d^{ABC}(T^{A})^i_j W^{B,\,\gamma} W^{C}_{\gamma}/32\pi^2$, and $\Lambda_{YM}\ra \Lambda_{SYM}.  $\label{(f7)}
}
\\

Let us introduce now into this Abelian  $U(1)^{N_c-1}$ dual gauge theory the large size $L$ Wilson loop with the quark-antiquark pair of external heavy test electric quarks. Each quark $Q_{\beta},\,\,\beta=1,...,N_c$ is characterized then by its electric charge vector  ${\vec E}_{\beta}(Q)$ with $N_c-1$ components in the base of $N_c-1$ independent standard Cartan charges. And similarly, each of $N_c-1$ {\it condensed} magnetic monopoles $M_n,\,\, n=1,...,N_c-1$ is characterized by its magnetic charge vector ${\vec M}_n({\rm Mon})$ with $N_c-1$ components. All quarks are mutually local between themselves. And $N_c-1$ magnetic  monopoles are mutually local between themselves. But electrically charged quarks and magnetic monopoles are, in general, mutually non-local. Their proper charge vectors satisfy the standard Dirac charge quantization conditions
\bq
{\vec E}_{\beta}(Q){\vec M}_n({\rm Mon})=Z_{\beta,n}/2\,,\quad \beta=1,...,N_c\,,\,\, n=1,...,N_c-1\,, \label{(5.3)}
\eq
where $Z_{\beta,n}$ is the integer number (including zero), depending on the choice of $\beta$ and $n$. And the vector ${\vec F}_{\beta}(Q)$ of the
total electric flux of $N_c-1$ Abelian electric photons emanating from electrically charged quark $Q_{\beta}$ is ${\vec F}_{\beta}(Q)=
{\vec E}_{\beta}(Q)$.\\

In the standard pure $SU(N_c)$ YM, when these $N_c-1$ magnetic monopoles are condensed in the vacuum state, then $N_c-1$ corresponding dual magnetic photons acquire masses $\mu_{\rm magn}\sim {\tilde g} \rho_{\rm magn}\sim {\tilde g}\Lambda_{YM}$. But condensation of magnetic  monopoles can not screen the fluxes ${\vec F}_{\beta}(Q)$ of electric photons emanating from heavy test quarks $Q_{\beta}$. As a result, these constant electric fluxes ${\vec F}_{\beta}(Q)={\vec E}_{\beta}(Q)$  are concentrated then in the confining tube of radius $R^{\perp}_{\rm magn}(Q)\sim \mu^{-1}_{\rm magn}$ connecting quark and antiquark at the distance $L\gg R^{\perp}_{\rm magn}$ between them. {\it The quantization conditions \eqref{(5.3)} are necessary for self-consistency and stability of these confining tubes}.\\

The properties of the Stuckelberg theory \cite{FS} are  qualitatively different. All $N_c^2-1$ electric gluons have now {\it non-dynamical but fixed by hands masses} $g|v| > 0$ in the bare perturbative Lagrangian, even at $g|v|\ll\Lambda_{\rm YM}$. For this reason, if there were confinement,  all total electric flux vectors ${\vec F}_{\beta}(Q,l)$ emanating along the tube from these test (anti)quarks, and so the (anti)quark corresponding electric charge vectors ${\vec E}_{\beta}(Q,l)= {\vec F}_{\beta}(Q,l)$, are screened now and their numerical values become dependent on the distance $l$ along the tube from the sources. And they all become exponentially small at distances $l > (g|v|)^{-1}$ along the tube from the sources. {\it This dependence on $l$ at $g|v|\neq 0$ of all $N_c$ charge vectors ${\vec E}_{\beta}(Q,l)$ breaks the charge quantization conditions \eqref{(5.3)} which were necessary for the self-consistency and stability} of the confining electric tube connecting quark and antiquark.\\

In other words, the normal self-consistent stable tubes with  the constant electric charge vectors ${\vec E}_{\beta}(Q)$ \eqref{(5.3)} along the tube, originating from $\mu_{\rm magn}\neq 0$ due to condensation of magnetic  monopoles, {\it can not be formed now at} $g|v| > 0$. But varying total electric quark flux ${\vec F}_{\beta}(Q,l)={\vec E}_{\beta}(Q,l)$ along the tube still can not be screened by these condensed magnetic  monopoles. As a result, $\mu_{\rm magn}$ {\it has to be zero} at fixed by hands $g|v|\neq 0$ for self-consistency, i.e. these normal dynamical magnetic  monopoles {\it can not condense} now in vacuum and produce nonzero masses $\mu_{\rm magn}\neq 0$ of corresponding dual magnetic photons, fixed $g|v|\neq 0$ prevents this condensation. This agrees with the statement in \cite{ch21} that {\it mutually non-local normal dynamical magnetic  monopoles and normal dynamical electric quarks can not condense simultaneously in the vacuum state}. As a result, {\it at all $0 < g|v| <\infty$, there is not confinement but screening of all electric charges in the theory} considered in \cite{FS}. And, as a result, there is not the phase transition but crossover between regions with  $g|v|\ll\Lambda_{\rm YM}$ and  $g|v|\gg\Lambda_{\rm YM}$ in the theory considered by E. Fradkin and S.H. Shenker in \cite{FS}.\\

A widely spread opinion (supported by lattice calculations) is that, in non-supersymmetric $SU(N_c)$ QCD (or in ${\cal N}=1$ SQCD) with {\it normal dynamical scalar quarks} $\phi^i_{\beta},\, i=1,...,N_F,\, \beta=1,...,N_c$, with all $2 N_F N_c$ their physical real degrees of freedom and with sufficiently large  bare and/or dynamical masses, the confinement of such non-higgsed quarks originates from higgsing (i.e. condensation) of $N_c-1$ magnetic  monopoles of $SU(N_c)$ QCD, with the typical magnetic condensate $\rho_{\rm magn}\sim \Lambda_{YM}\sim\Lambda_{QCD}$. But such higgsed magnetic  monopoles, ensuring confinement in YM, are mutually nonlocal with these electrically charged quarks. For this reason, such magnetic  monopoles with the much larger vacuum condensate  $\rho_{\rm magn}\sim\Lambda_{YM}\sim\Lambda_{QCD}\gg g|v|$ {\it will keep normal dynamical scalar electric quarks} $\phi^i_{\beta}$ {\it confined and will prevent them from condensing with} $|v|\neq 0$ {\it in the vacuum state}. This allows then to form the normal self-consistent and stable confining tubes with constant electric charge vectors ${\vec E}_{\beta}(Q)$ \eqref{(5.3)}.\\

The normal dynamical particles with larger vacuum condensate overwhelm. E.g., the normal scalar magnetic monopoles with the vacuum condensate $\rho_{\rm magn} > \rho_{\rm electr}$ will be really condensed in vacuum with $\rho_{\rm magn}\neq 0$, keep normal electrically charged scalar quarks mutually non-local with them confined and prevent these last from condensing in vacuum, i.e. $\rho_{\rm electr}=0$ then. And vice versa at $\rho_{\rm magn} < \rho_{\rm electr}$.
\footnote{\,
In the standard non-supersymmetric $SU(N_c),\,\,N_F=N_c$ QCD with normal dynamical light scalar quarks $\phi^i_{\beta}$ with all $2 N_c^2$ their
physical real degrees of freedom, these quarks are not massless even at $m_{\phi}= 0$ and zero self-interaction potential in the bare perturbative Lagrangian. Even at small $|v|\ll\Lambda_{QCD}$ (where $|v|$ is e.g. a {\it potentially possible} dynamical quark condensate, $\langle{\hat\phi}^i_{\beta}\rangle=\delta^i_{\beta}|v|\,$), they acquire in this case non-perturbative dynamical masses $\mu^2_{\phi}\sim  V_{HQ}\sim \Lambda^2_{QCD}\sim\Lambda^2_{YM},\,\,\,\frac{1}{N_c}\sum_{\beta=1}^{N_c}\langle (\phi^\dagger)_j^{\beta}\phi^i_{\beta}\rangle_{HQ} \sim\delta^i_j\,V_{HQ}$. And, connected with this, at $|v| \ll\rho_{\rm magn}\sim\Lambda_{YM}\sim\Lambda_{QCD}$ {\it there is confinement of all electrically charged particles with the string tension} $\sigma^{1/2}\sim\rho_{\rm magn}\sim\Lambda_{YM}\sim\Lambda_{QCD}$. So that, $\sigma^{1/2}\sim\rho_{\rm magn}\sim\Lambda_{QCD}\sim\Lambda_{YM}\gg |v|$ will keep really such standard quarks confined and will prevent them from condensing with $|v|\neq 0$ in vacuum.  Such {\it standard  quarks will be really in the HQ(heavy quark)-phase at $|v|\ll\Lambda_{QCD}$, i.e. not higgsed but confined} in this case, with $\langle{\hat\phi}^i_{\beta}\rangle_{HQ}=\delta^i_{\beta} |v|_{HQ}=0$ really. And  $V_{HQ}\sim\Lambda^2_{YM}$ is {\it non-factorizable} for such quarks in the HQ-phase with $\mu_{\phi}\sim [\,V_{HQ}\,]^{1/2}\sim\Lambda_{QCD}\neq |v|_{HQ}=0$. This is qualitatively similar to section 4.1 in \cite{ch21} with standard massive confined quarks in the HQ (heavy quark)-phase.

Standard  scalar quarks $\phi^i_{\beta}$ with  $m_{\phi}= 0$ and zero self-interaction quark potential in the bare perturbative Lagrangian will be really higgsed in the weak coupling regime at e.g. $\langle {\hat\phi}^i_{\beta}\rangle_{higgs}=\delta^i_ {\beta}|v|_{\rm higgs},\,\,\,i,\beta=1,...,N_F=N_c$, with $g|v|_{\rm higgs}\gg\Lambda_{QCD}$. All $N_c^2-1$ electric gluons will acquire then large masses $\sim g|v|_{\rm higgs}\gg\Lambda_{QCD}$ in this case, because $V_{higgs}\sim |v|_{\rm higgs}^2\gg\Lambda^2_{QCD}$ will be {\it factorized} now. $N_c^2$ quark real degrees of freedom will also acquire dynamical masses $\sim g|v|_{\rm higgs}\gg\Lambda_{QCD}$ and form one adjoint and one singlet representations of $SU(N_F=N_c)$. And there will be {\it the phase transition} between the regions with values of the gauge invariant order parameter $\rho=|v|_{\rm higgs}\gg\Lambda_{QCD}$ in the higgs phase and $\rho=|v|_{HQ}=0$ in the confinement phase. This is similar to the phase transitions in \cite{ch21}.

While in ${\cal N}=1$ SQCD, non-higgsed quarks have $m^{\rm pole}_Q\ra 0$ at $m_Q\ra 0$ \eqref{(2.7)},\eqref{(2.1.1)} (or $m^{\rm pole}_{H}\ra 0$ at $m_H\ra 0$ \eqref{(2.2.6)},\eqref{(2.2.10)}), and the tension of confining string is $\sigma^{1/2}\sim\lym\ra 0$ in the limit $\mi\ra 0$ with fixed $r=m_L/m_H$. And so, there is no confinement, even if the chiral symmetry of the SUSY Lagrangian is broken spontaneously, see \eqref{(2.3)}-\eqref{(2.5)},\eqref{(2.2.1)}. \label{(f8)}
}
\\

And the model described above illustrates explicitly the mechanism underlying the statement in \cite{ch21} that mutually non-local normal magnetic monopoles and normal scalar quarks with all  their physical degrees of freedom {\it can not be condensed simultaneously in the vacuum state}.\\

Let us make now very short remarks about our ordinary (i.e. not SUSY) $SU(3)_c$ color QCD with three massless  fermionic quarks in this model of confinement.
\footnote{\,
Only color quantum numbers are shown explicitly below. The flavor and Lorentz quantum numbers  are not shown.
}
The condensed $\langle ({\hat H}^{\rm adj})^i_j(0)\rangle$ breaks the color $SU(3)_c\ra U^{2}(1)$, see \eqref{(5.2)}. All three colored gauge invariant quark fields $\hat{\psi}_1, \hat{\psi}_2, \hat{\psi}_3$  {\it are confined}   ( see section 2 in \cite{ch21} for the meaning of the hat).  But, by themselves,  the gauge invariant composite fields ${\hat C}_0=({\hat{\ov \psi}}_1 {\hat \psi}_1+{\hat{\ov \psi}}_2 {\hat\psi}_2+{\hat{\ov \psi}}_3 {\hat \psi}_3),\,\,{\hat C}_3=({\hat{\ov \psi}}_1 {\hat \psi}_1-{\hat{\ov \psi}}_2 {\hat\psi}_2),\,\,{\hat C}_8=({\hat{\ov \psi}}_1 {\hat \psi}_1+{\hat{\ov \psi}}_2 {\hat\psi}_2-2\,{\hat{\ov \psi}}_3 {\hat \psi}_3)$ do not emit electric Cartan color fluxes and so are {\it not confined}. 

On account of spontaneous breaking of global chiral $SU(3)_L\times SU(3)_R$ symmetry, massless quarks acquire equal dynamical (constituent) masses $\sim\Lambda_{QCD}\sim\Lambda_{YM}$.

As a result,  ignoring possible mixings, there are e.g. three lowest lying massive $\rho$-meson states  $\rho_{0},\, \rho_{3},\, \rho_{8}$ in spectral densities of ${\hat C}_0,\, {\hat C}_3,\, {\hat C}_8$. 
\footnote{\,
But we expect naturally that lowest massive states  in spectral densities of ${\hat C}_3$ and ${\hat C}_8$  are noticeably heavier than  the corresponding lowest massive state in the spectral density of  ${\hat C}_0$.
}
(The exceptions for possible mixings are 8 massless genuine flavored Goldstone bosons in the
spectral density of pseudoscalar $C_0,\,\,\,\pi_{0.i},\,i=1...8$). 

Therefore, the hadron quantum numbers of the standard quark model have to be reconsidered. 

More clearly all these spectral properties will be seen in spectra of charmonia and especially bottomonia.\\

In support of the conclusion of E. Fradkin and S.H. Shenker \cite{FS} about the crossover between the confinement and higgs regimes, it was written by K. Intriligator and N. Seiberg in \cite{IS} for ${\cal N}=1\,SU(N_c)$ SQCD with $N_F=N_c$ the following (for the moduli space, with $m_{Q,i}\ra 0)$.\, -

''For large expectation values of the fields,
\footnote{\,
 the values of $\langle M^i_i\rangle$ are implied  \label{(f9)}
 }
a Higgs description is most natural while, for small expectation values, it is more natural to interpret the theory as 'confining'...
\footnote{\,
the quotes are set here because the string between heavy test quark and antiquark breaks with increased distance between them due to the birth of a light dynamical quark-antiquark pair from the vacuum
}
Because these theories (i.e. ${\cal N}=1$ SUSY QCD) have matter fields in the fundamental representation of the gauge group, as mentioned in the introduction, there is no invariant distinction between the Higgs and the confining phases \cite{FS}. It is possible to smoothly interpolate from one interpretation to the other''.

In other words. {\it Because one can move completely smoothly (i.e. analytically)}  on the moduli space from e.g. the region  {\bf A}  with $\langle M_L\rangle\gg\la^2,\, \langle M_H\rangle\ll\la^2$, see \eqref{(2.2.1)}, where LL-quarks $Q^L_L, {\ov Q}^L_L$ are higgsed while (according to \cite{IS}) HH-quarks $Q^H_H,\, {\ov Q}^H_H$ are confined, to another region {\bf B} with $\langle M_L\rangle\ll\la^2,\, \langle M_H\rangle\gg\la^2$ where, vice versa, quarks $Q^H_H, {\ov Q}^H_H$ are higgsed while quarks $Q^L_L,\, {\ov Q}^L_L$  are confined, this is the independent confirmation of the conclusions of paper \cite{FS} that there is the analytic crossover between the confinement and higgs regimes, not the non-analytic phase transition.\\

There are two loopholes in these arguments. It is right that quarks $Q^L_L, {\ov Q}^L_L$ or $Q^H_H, {\ov Q}^H_H$ are higgsed on the moduli space in regions {\bf A} or {\bf B} with $\langle M_L\rangle\gg\la^2$ or $\langle M_H\rangle\gg\la^2$ respectively (at not too large $N_c$, see footnote \ref{(f5)}). But first, as emphasized above in this section,  the quarks with $\langle M_{L,H}\rangle\ll\la^2$ are not confined.  {\it There are regions on the moduli space where some quarks are higgsed, but there are no regions where some quarks are confined. There is no confinement on the moduli space of the direct theory because the string tension $\sigma^{1/2}\sim\lym\ra 0$} at $\mi\ra 0$, see \eqref{(2.3)},\eqref{(5.1)} and footnote \ref{(f1)}. Therefore, unlike the statements in \cite{IS}, the transitions between the regions {\bf A} and {\bf B} of the moduli space are not the transitions between the higgs and confinement regimes. These are transitions between regimes of higgsed or not higgsed some quarks, but in all regions all quarks are not confined.\\

Second, let us start on the moduli space, i.e. at $m_{L,H}\ra 0$ with fixed $0 < r=m_L/m_H < \infty$ in the region {\bf A} and move to the region {\bf B}. In the region {\bf A} at $\mu_{\rm gl, L}\gg\la$, LL-quarks are higgsed, see \eqref{(2.2.1)},\eqref{(2.2.2)},  while HH-quarks are {\it not higgsed and not confined} (because $\sigma^{1/2}\sim\lym\ra 0$ at $\mi\ra 0$). In the region {\bf B} at $\mu_{\rm gl, H}\gg\la$, vice versa,  HH-quarks are higgsed while {\it LL-quarks are not higgsed and not confined}.

On the way between regions {\bf A} and {\bf B}\,, there is the point $r=1$ with equal mass quarks and 
$\langle M^i_i\rangle=\la^2$. As shown in \eqref{(2.7)} and in section 2.1, all quarks are {\it not higgsed} at this point at $m_Q > 0$, even at not large $N_c$.  And this remains so even in the chiral limit $m_Q\ra 0$. In this limit, the whole mass spectrum consists of massless quarks and gluons, see the end of section 2.1. All particles are {\it not higgsed and not confined} (because $\lym\ra 0$).  Although $\langle M^i_i\rangle=\la^2$, as it follows from the Konishi anomaly \eqref{(2.4)}, but it is {\it non-factorizable}, i.e $\langle M^i_i\rangle=\la^2\neq \sum_{\beta=1}^{N_c}\langle{\ov Q}^{\beta}_i\rangle\langle Q^i_{\beta}\rangle=0$ at $r=1$.\\

In section 2 of \cite{ch21} the {\it colored} gauge invariant order parameter was introduced. It looks as:\,  $\langle {\hat Q}^i_{\beta}
\rangle=\delta^i_{\beta} \rho,\,\, \langle{\hat {\ov Q}}^{\,{\beta}}_i\rangle=\delta^{\beta}_i \rho,\,\,i=1,...,N_F,\,\beta=1,...,N_c$, where
${\hat Q}^i_{\beta}$ is the {\it colored but gauge invariant quark field}.   The gauge invariant order parameter $\rho$ behaves {{\it non-analytically}. It is nonzero for higgsed quarks. While it is zero if quarks are not higgsed, independently of whether they are confined or not}, see \cite{ch21} for all details.

At  fixed $N_c$ and $0 < r=m_L/m_H\ll 1$, such that $\langle M_L\rangle\gg\la^2,\, \mu^2_{\rm gl,L}\gg\la^2$ \eqref{(2.2.1)},\eqref{(2.2.2)}, LL-quarks are definitely higgsed, while HH-quarks with $\langle M_H\rangle\ll\la^2$ are definitely not higgsed (because they are not higgsed even at $\langle M_H\rangle=\la^2$).  And similarly,  at $r\gg 1$ HH-quarks are higgsed, while LL-quarks are not.

As it is seen from \eqref{(2.2.1)},\eqref{(2.2.2)}, at some fixed $N_c$, to have definitely higgsed LL-quarks, $r$ has to be sufficiently small:\, $r < r_{L}(N_c)$, where $r_{L}(N_c)$ is sufficiently smaller than unity and is such that $\mu_{\rm gl,L}$ is sufficiently larger than $\la$. And similarly for HH-quarks. To have definitely higgsed HH-quarks, $r$ has to be sufficiently large:\,  $r > r_H(N_c)$, where  $r_{H}(N_c)$ is sufficiently larger than unity and is such that $\mu_{\rm gl,H}$ is sufficiently larger than $\la$.  Therefore, at fixed $N_c$, there are {\it two phase transitions at the points $r=r_L(N_c)$ and  $r=r_H(N_c)$ on the way from the region {\bf A} to {\bf B}} where, respectively, the LL-quarks become unhiggsed and HH-quarks become higgsed. And there is the finite width region $r_L(N_c) < r < r_H(N_c)$ where all quarks are not higgsed. The gauge invariant order parameter $\rho$ \cite{ch21} is nonzero outside the region  $r_L(N_c) < r < r_H(N_c),\,\,\rho_{\rm higgs}\neq 0$, while it is zero inside it, $\rho_{HQ}=0$.

So that, the arguments of K. Intriligator and N. Seiberg in \cite{IS}, put forward in support of the crossover \cite{FS}, about an analytical movement along the moduli space and an absence of phase transitions are erroneous.

\vspace*{- 4mm}

\section{Conclusions}

1)\,\, As shown in sections 2 and 3,  the mass spectra of particles with masses $<\la$ of the direct $SU(N_c),\,\,\, N_F=N_c$ $\,\,{\cal N}=1$ SQCD theory  and of Seiberg's dual one \cite{S1} are parametrically different, both for equal or unequal quark masses $m_{Q,i}\neq 0$. Therefore, these two theories are not equivalent.

So that, the proposal by N. Seiberg \cite{S1} of his non-gauge $N_F=N_c$ dual ${\cal N}=1$ SQCD theory with only mesons $M^i_j$ and baryons $B,\,{\ov B}$ (with one constraint) as the genuine low energy form of the direct $SU(N_c)\,\, N_F=N_c$ ${\cal N}=1$ SQCD theory due to the strong confinement with the string tension $\sigma^{1/2}\sim\la$ (the so called "S-confinement") is erroneous. And similarly for the case $N_F=N_c+1$, see Appendix "B" in \cite{Session}. As for differences in mass spectra of the direct and Seiberg's dual ${\cal N}=1$ SQCD or ${\cal N}=1$ SQCD-type theories at other values  $N_F > N_c+1$, see e.g. \cite{ch3},\cite{ch5},\cite{Session}.

2)\,\, E. Fradkin and  S.H. Shenker considered in \cite{FS} the very special non-supersymmetric lattice $SU(N_c)$  QCD theory with $N_F=N_c$ scalar
bifundamental quarks $\Phi^i_{\beta},\,\,i=1...N_F=N_c,\,\,\beta=1...N_c$. In the unitary gauge, from $2 N_c^2$ physical real degrees of freedom of these quarks $\Phi^i_{\beta}$, were retained only $N_c^2-1$ real Goldstone modes eaten by gluons. All other $N_c^2+1$ real physical degrees of freedom were deleted by hands: $\Phi^i_{\beta}= \delta^i_{\beta}|v|,\,\,|v|={\rm const > 0}$. The conclusion of \cite{FS} was that the transition between the confinement regime at small $0 < |v|\ll\Lambda_{QCD}$ (according to \cite{FS}) and the higgs regime at large $|v|\gg\Lambda_{QCD}$ is the analytic crossover, not the non-analytic phase transition.

This model used in \cite{FS} is criticized in section 5 of this paper (see also page 9 in v6 of \cite{ch21}\,) as {\it incompatible} with and qualitatively different from the standard $SU(N_c),\,\,N_F=N_c$ QCD theory with standard scalar quarks $\phi^i_{\beta}$ with all $2 N_c^2$ their physical real degrees of freedom, see footnote \ref{(f8)}. Really, this theory \cite{FS} is the Stueckelberg {\it $SU(N_c)$ YM theory with no dynamical quarks} and with massive all $N_c^2-1$ electric gluons with fixed by hands masses $g|v| > 0$. In this theory, in any case, at large distances, the total electric flux emanating e.g. from external heavy test (anti)quark is exponentially screened due to fixed nonzero masses of all electric gluons and {\it there is no genuin confinement} at all $|v|\neq 0$. In other words, this theory \cite{FS} with such defective "quarks"\, stays really permanently in the completely higgsed phase at all $|v|\neq 0$, and this is a reason for the crossover between regions  $0 < |v|\ll\Lambda_{QCD}$ and  $ |v|\gg\Lambda_{QCD}$
in \cite{FS}.\\

Besides, the arguments presented in \cite{IS} by K. Intriligator and N. Seiberg for the standard direct $SU(N_c),\,\, N_F=N_c$ $\,\,\,{\cal N}=1$ SQCD theory in support of conclusions \cite{FS} about crossover are criticized in section 5 as erroneous.

\end{document}